\documentclass[10pt,
floatfix, 
aps, 
pra, 
amsmath,
twocolumn,
superscriptaddress
]{revtex4-2}

\pdfoutput=1

\usepackage{graphicx}
\usepackage{amsmath}
\usepackage{amssymb}
\usepackage{bbm}
\usepackage[hidelinks]{hyperref}
\usepackage{soul}
\usepackage{xcolor}
\usepackage{bbold}
\usepackage{physics}
\usepackage{mathtools}

\newcommand{\dket}[1]{\ket{\overline{#1}}}
\newcommand{\dbra}[1]{\bra{\overline{#1}}}
\newcommand{\dmel}[2]{\langle \overline{#1}\ket{\overline{#2}}}

\begin{document}

\title{Measurement-induced state transitions in dispersive qubit readout schemes}

\author{Konstantin N. Nesterov}
\email{konstantin@atlantic-quantum.com}
\affiliation{Atlantic Quantum, Cambridge, MA 02139, USA}

\author{Ivan V. Pechenezhskiy}
\email{ivpechen@syr.edu}
\affiliation{Department of Physics, Syracuse University, Syracuse, NY 13244, USA}
\date{\today}

\begin{abstract}
The dispersive-readout scheme enables quantum nondemolition measurement of superconducting qubits. An increased readout power can shorten the readout time and reduce the state discrimination error but can promote qubit transitions into higher noncomputational states. The ability to predict the onset of these measurement-induced state transitions can aid the optimization of qubit circuits and provide means for comparing the readout performance of different qubit types. Building upon the concept of dressed coherent states, we consider two straightforward metrics for determining the maximum number of photons that can be used for dispersive readout without causing state transitions. We focus on the fluxonium readout to demonstrate the independence of the metrics from any qubit-type-specific approximations. The dispersive readout of transmons and other superconducting qubits can be treated universally in the same fashion.

\end{abstract}

\maketitle

\section{Introduction}

The dispersive readout of a superconducting qubit is performed by probing the qubit-state-dependent frequency of a~linear resonator coupled to the qubit~\cite{Blais04, Blais21}, which enables a fast and high-fidelity single-shot measurement~\cite{Johnson2012, Walter2017, Swiadek2024}. A short measurement duration is vital for implementing error-correction codes such as the surface code~\cite{Fowler2012}, in which the data qubits must idle during the readout of the measure qubits and, thus, unavoidably accumulate errors due to intrinsic lifetime limitations~\cite{Acharya23}. Importantly, preserving the quantum nondemolitionness of the dispersive readout would guarantee that a~qubit remains in a computational state after the measurement, allowing straightforward reset protocols~\cite{Magnard2018, Sunada2022}.

Several parameters determine the measurement rate, with the discussions typically framed in terms of the resonator dispersive shift~$\chi$, the photon decay rate $\kappa$, and the average photon number~$\bar{n}$. Numerical simulations can easily predict the first two quantities for a~given qubit design. The optimal readout condition is often stated as $\kappa\approx 2\chi$, which maximizes the signal-to-noise ratio (SNR) for a fixed~$\bar{n}$ in a linear resonator~\cite{Gambetta2008, Boissonneault2010}. Naïvely, increasing $\bar{n}$ should directly result in a higher SNR and thus in a better readout~\cite{Gambetta2008}. Unfortunately, in addition to a stronger resonator nonlinearity at larger $\bar{n}$, the qubit undergoes measurement-induced state transitions (MISTs) when the number of photons exceeds a~threshold value~\cite{Khezri23, Sank2016, Shillito2022, Dumas2024}. These transitions, sometimes referred to as qubit ionization in the context of transmon qubits escaping the Josephson potential well~\cite{Shillito2022, Dumas2024, Lescanne2019, Nojiri2024}, limit the measurement rate, degrade readout fidelity, and complicate qubit reset.

In this paper, we theoretically investigate the onset of MISTs during the dispersive readout of superconducting qubits by focusing on two metrics that exhibit clear signatures of MISTs. The first metric, qubit purity, quantifies the qubit-resonator entanglement and was used in the past in numerical studies of qubit transitions and structural instabilities of the system dynamics~\cite{Verney2019, Shillito2022}. Time-domain simulations reveal that the qubit purity remains very close to unity unless a state transition occurs as the number of readout photons increases. Our explanation of this sensitivity is based on the notion that away from MISTs, the state of a driven qubit-resonator system would be close to a dressed coherent state~\cite{Sete2013, Sete2014, Govia2016, Khezri2016}. Perhaps surprisingly, such a state and its squeezed modifications have been shown to be almost unentangled~\cite{Khezri2016}. Further exploring this picture, we propose another metric that characterizes deviations in the drive matrix elements computed for the dressed coherent states. This metric, which is also very sensitive to MISTs, bypasses time-domain simulations and only requires accurate identification of the interacting qubit-resonator states. Throughout the paper, we focus on the case of a~fluxonium qubit capacitively coupled to a resonator, although the analysis applies to the dispersive readout of any qubit type with a limited number of internal degrees of freedom. 

The Jaynes-Cummings Hamiltonian~\cite{Jaynes63} provides the starting point for the readout analysis of an ideal two-level qubit coupled to a resonator. In the dispersive regime, when the qubit-resonator coupling $g$ is much smaller than the qubit-resonator detuning $\Delta$, the resonator frequency is pulled by approximately $\pm g^2/\Delta$ with the sign determined by the qubit state~\cite{Blais04}. The dispersive approximation breaks down when the photon number becomes comparable to $n_\textrm{crit}=\Delta^2/(4g^2)$, which corresponds to significant hybridization of the bare qubit-resonator states and the onset of resonator nonlinearity~\cite{Boissonneault2010}. Although $n_\textrm{crit}$ is often quoted as a characteristic of a qubit-resonator system, it does not set a hard limit on resonator occupation in a~readout even when generalized for realistic multi-level qubits~\cite{Dumas2024}. In particular, $n_\textrm{crit}$ does not predict the likelihood of MISTs, which, depending on the system parameters, may occur at $\bar{n}$ drastically different from $n_{\rm crit}$. MISTs in dispersive readout schemes are poorly understood and have been studied only in the context of transmon qubits as summarized below~\cite{Sank2016, Khezri23, Qi2019, Shillito2022, Cohen2023, Xiao2023, Dumas2024, Hirasaki2024}. 

In a nutshell, the explanation for MISTs in transmons is based on the resonances between energy levels of the interacting qubit-resonator system corresponding to different transmon states. The rotating-wave approximation (RWA) simplifies the analysis for this qubit type, 
permitting an intuitive picture of level resonances based on so-called RWA strips. Each RWA strip is defined as the group of levels in the composite qubit-resonator Hilbert space with a fixed total number of qubit and resonator excitations~\cite{Sank2016}. There are two distinct parameter regimes determined by the transmon and resonator frequencies $\omega_q$ and~$\omega_r$. When $\omega_q>\omega_r$, the energies of the levels within the same RWA strip are not monotonic with the qubit index due to the negative transmon anharmonicity. This is visually represented as the bending of the strip over itself~\cite{Khezri23}, which can result in a resonance between one of the two lowest qubit levels and a level near the edge of the transmon's cosine potential, causing MISTs~\cite{Khezri23}. The onset of MISTs often happens at relatively small photon numbers, depends strongly on the qubit-resonator detuning $\Delta$, and is highly sensitive to the transmon offset charge $n_g$ because of an increased sensitivity of the higher-lying qubit levels to~$n_g$. The experimental data is consistent with a~model based on the semiclassical equations of motions for the resonator field and the effective Schr\"odinger equation for the qubit~\cite{Khezri23}.

In the opposite regime, when $\omega_q < \omega_r$, the resonances between levels in different RWA strips are found to be responsible for the transitions~\cite{Sank2016}. Crucially, while a~simple model based on diagonalization of the Hamiltonian in the RWA is sufficient to identify the resonances, non-RWA terms are necessary to explain the presence of the transitions between different RWA strips, when the total number of excitations is not conserved. This points to the potential limitations of overly simplified models in predicting MISTs. Notably, the transitions were observed at $\bar{n}$ several times greater than $n_{\rm crit}$, confirming that $n_{\rm crit}$ is not a~reliable metric for estimating the limits on the dispersive readout power. In the $\omega_q<\omega_r$ regime, temporary resonance conditions due to fluctuations in $n_g$ may cause MISTs only during specific time intervals~\cite{Hirasaki2024}.

A computationally expensive numerical study of the full dissipative dynamics of a transmon-resonator system under a~strong measurement drive also revealed signatures of qubit transitions~\cite{Shillito2022}. In that study, the Lindblad master equation with non-RWA terms in the Hamiltonian was integrated using large-scale computational accelerators, while a~semiclassical approach was used to interpret the results. Similarly to Refs.~\cite{Sank2016, Khezri23}, qubit transitions were attributed to level resonances occurring at specific photon numbers. The photon numbers were found to be strongly parameter dependent and sometimes very small. Theoretical understanding of MISTs in transmons was further advanced through a comparison of the fully quantized model to the Floquet analysis and an entirely classical model of a driven nonlinear pendulum~\cite{Cohen2023, Dumas2024}.

The onset of MISTs in the other superconducting qubit types remains unexplored in part because of the inapplicability of the RWA-based approaches, such as in the case of fluxonia~\cite{Manucharyan09, Nguyen2019, Nguyen2022}. However, as with transmons, the dispersive shifts in fluxonium-resonator systems can be straightforwardly predicted with simulations matching experiments~\cite{Zhu2013a}, and large photon numbers have been experimentally used for reading out fluxonia~\cite{Gusenkova2021}. Yet, we are unaware of any practical tools for predicting the onset of MISTs in fluxonia with an increase of $\bar{n}$.

Here, we investigate generic approaches for predicting MISTs in superconducting qubits with a known Hamiltonian model using fluxonium as an example. The outline of this paper is as follows. In Sec.~\ref{Sec-model}, we introduce the Hamiltonian model, describe the procedure of numerical identification of its eigenstates, and discuss a~particular example of a fluxonium spectrum prone to qubit transitions. In Sec.~\ref{Sec-metrics}, we simulate the dynamics for this specific example, discuss state transitions, and introduce metrics to catch the MIST effects. In Sec.~\ref{Sec-dependence-on-parameters}, we investigate the dependence of MISTs on the external magnetic flux and the resonator frequency. We conclude in Sec.~\ref{Sec-conclusion} with general remarks. In Appendix~\ref{appendix:dispersive}, we discuss the formal breakdown of the dispersive approximation and calculate the dispersive critical photon number, while in Appendix~\ref{appendix:state_identification}, we discuss the step-size selection for the state-identification algorithm. In Appendix~\ref{appendix:purity}, we take a~closer look at the qubit purity by calculating it in several limits. In Appendix~\ref{appendix:dissipation}, we simulate the qubit purity in the presence of resonator decay. In Appendix~\ref{appendix:perturbation_mat_el}, the matrix element of the raising operator $\hat{a}^\dagger$ is calculated using the perturbation theory. In Appendix~\ref{appendix:comparison}, we compare the metrics in more detail. In Appendix~\ref{appendix:transmon}, we provide the simulation results for a~transmon using parameters from Ref.~\cite{Khezri23} to further justify our approach.

\section{Model}\label{Sec-model}

\subsection{Hamiltonian}

\begin{figure}
 \includegraphics[width=.76\columnwidth]{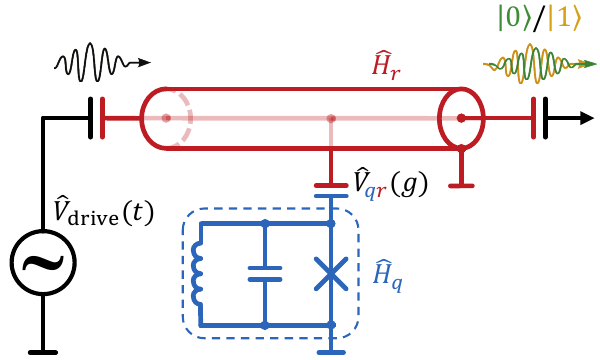}
 \caption{Schematic for dispersive readout of a superconducting qubit. A drive tone probes a readout resonator coupled to the qubit with coupling strength $g$. For illustrative purposes, we consider the case of capacitive coupling and fluxonium qubit, although the analysis can be generalized to inductive coupling and other superconducting qubit types.}\label{fig:diagram}
 \vspace{-15pt}
\end{figure}

We propose a universal method for identifying the onset of MISTs in a general class of circuit quantum electrodynamics (cQED) setups (Fig.~\ref{fig:diagram}) described by the Hamiltonian
\begin{equation}
\hat{H} = \hat{H}_q + \hat{H}_r+ \hat{V}_{qr}(g) + \hat{V}_\textrm{drive}(t)\,.
\label{eq:model}
\end{equation}
Here, $\hat{H}_q$ is a qubit Hamiltonian. In the main text, we examine a fluxonium qubit~\cite{Manucharyan09}, for which
\begin{equation}\label{eq:H_fluxonium}
    \hat{H}_q = 4E_C \hat{n}^2 + E_L \hat{\varphi}^2/2 - E_J \cos(\hat{\varphi} - \varphi_{\textrm{ext}})\,,
\end{equation}
while in Appendix~\ref{appendix:transmon} we include results for a~transmon~$\hat{H}_q$. In Eq.~\eqref{eq:H_fluxonium}, $\hat{\varphi}$ and $\hat{n}$ are the normalized flux and charge operators satisfying $[\hat{\varphi}, \hat{n}] = i$, and $E_C$, $E_J$, and $E_L$ are the charging, Josephson, and inductive energies.  In addition, $\varphi_{\textrm{ext}}=2\pi\Phi_\mathrm{ext}/\Phi_0$, where $\Phi_{\textrm{ext}}$ is the external flux through the superconducting loop, $\Phi_0=h/(2e)$ is the flux quantum, $-e$ is the electron charge, and $h=2\pi\hbar$ is the Planck constant. 

The second term in Eq.~\eqref{eq:model} describes the (linear) resonator and is written in terms of the raising and lowering operators $\hat{a}^\dagger$ and $\hat{a}$ as $\hat{H}_r=\hbar\omega_r \hat{a}^\dagger \hat{a}$, where $\omega_r$ is the bare resonator frequency. The third term, $\hat{V}_{qr}(g)$, describes the qubit-resonator coupling, which is parameterized by the coupling strength $g$. For a pure capacitive coupling, considered in this paper,
$\hat{V}_{qr}(g)=i\hbar g\hat{n}(\hat{a}^\dagger - \hat{a})$, while for pure inductive coupling $\hat{V}_{qr}(g)=\hbar g\hat{\varphi} (\hat{a}^\dagger + \hat{a})$. A~readout drive of a fixed drive frequency~$\omega_\textrm{drive}$ can be specified by 
\begin{equation}\label{eq:drive}
 \hat{V}_\textrm{drive}(t)=2i\hbar\varepsilon(t)(\hat{a}^\dagger - \hat{a})\cos(\omega_\textrm{drive}t)\,, 
\end{equation}
where $\varepsilon(t)$ is the drive strength incorporating an initial raising stage. Generally, though, the time dependence of $\hat{V}_\textrm{drive}(t)$ can be more complex~\cite{McClure2016}, and the drive is not necessarily monochromatic.

\subsection{State identification}\label{sec-model-dasi}

Let $\hat{H}_{\rm idle}(g)$ be the Hamiltonian~\eqref{eq:model} without the drive term~$\hat{V}_\textrm{drive}(t)$. When $g=0$, its eigenstates are trivially $\ket{k, n} = \ket{k}\otimes\ket{n}$ with total energies $E_{k,n}=E_k+n\hbar\omega_r$, where $\ket{k}$ is the $k$-th eigenstate of the qubit Hamiltonian $\hat{H}_q$ with energy $E_k$ and $\ket{n}$ is the $n$-th eigenstate of the resonator Hamiltonian $\hat{H}_r$, $k, n \ge 0$. When $g\neq 0$, the bare states $\ket{k, n}$ are no longer eigenstates of $\hat{H}_{\rm idle}(g)$. For convenience, we use the same indices to label the dressed eigenstates of the interacting Hamiltonian and the corresponding eigenenergies as $\dket{k,n}_g$ and $E_{\overline{{k,n}}}(g)$, or simply ${\dket{k,n}}$ and $E_{\overline{k,n}}$ whenever we do not need to emphasize the value of $g$. 

Ideally, a particular eigenstate of $\hat{H}_{\rm idle}(g)$ should be identified as $\dket{k,n}_g$ when it is connected to $\ket{k, n}$ adiabatically, i.e., by varying $g$ slowly. In our numerical simulations, we use a labeling algorithm that we refer to as discrete adiabatic state identification (DASI). Starting with the noninteracting states $\ket{k,n}\equiv\dket{k,n}_{g=0}$, the coupling strength $g$ is gradually increased in small discrete increments $\delta g$  until the value of interest is reached. At each new step, the updated interacting Hamiltonian $\hat{H}_{\rm idle}(g+\delta g)$ is diagonalized, and its eigenstates $\dket{k, n}_{g+\delta g}$ are identified by maximizing the overlaps with $\dket{k, n}_g$, the eigenstates of $\hat{H}_{\rm idle}(g)$. We emphasize that as a~result, a state $\dket{k, n}_g$ can have the largest overlap with a~noninteracting state that is different from $\ket{k, n}$ due to nontrivial state hybridization. A simpler approach based on maximizing overlaps between dressed and bare states works well when only low-lying eigenstates are needed, such as when calculating dispersive shifts, but fails in general.

\begin{figure}
 \includegraphics[width=\columnwidth]{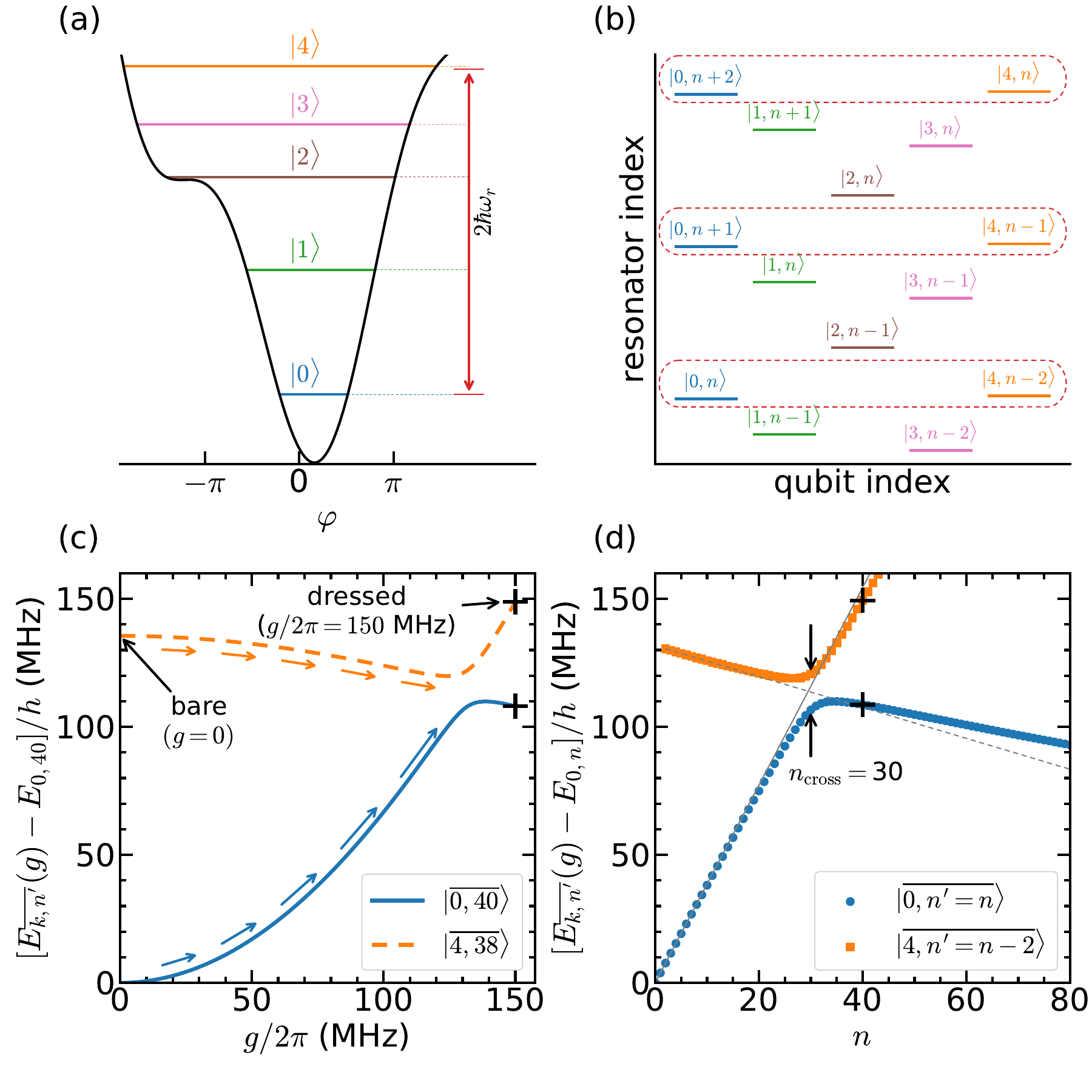}
 \caption{(a)~Potential-energy profile and single-excitation energy level diagram for a fluxonium qubit at $\Phi_\mathrm{ext}/\Phi_0=0.1$ with the other system parameters specified in the text. The doubled bare resonator frequency is approximately equal to the transition frequency between 
 states $\ket{0}$ and $\ket{4}$, leading to a~resonant condition at a certain photon number. (b)~Energy-level diagram of the combined qubit-resonator system with no interaction ($g=0$). (c)~Identification of the dressed eigenstates via a gradual sweep of the coupling strength~$g$ from zero (bare states) to $g/2\pi=150\,\mathrm{MHz}$ (black crosses). For clarity, only the dressed eigenenergies $E_{\overline{0,40}}$ and $E_{\overline{4,38}}$ relative to $E_{0,40}$ are shown. (d)~Dressed eigenenergies $E_{\overline{0,n'=n}}$ and $E_{\overline{4,n'=n-2}}$ versus photon number $n$ for $g/2\pi=150\,\mathrm{MHz}$. The energies are shifted by $E_{0,n}=n\hbar\omega_r$ to remove the linear contribution proportional to $n$. The avoided-level-like crossing of the eigenladders at $n_\textrm{cross}=30$ can lead to a MIST when a~comparable number of photons is used for readout. Gray lines show energies calculated in the dispersive approximation.}\label{fig:levels}
\end{figure}

Throughout the main text, we consider a fluxonium qubit with $E_J/h=4.0\,\mathrm{GHz}$, $E_C/h=1.0\,\mathrm{GHz}$, and $E_L/h=1.0\,\mathrm{GHz}$ capacitively coupled with coupling strength $g/2\pi=150\,\mathrm{MHz}$ to a resonator with $\omega_r/2\pi=7.0\,\mathrm{GHz}$. In our numerical simulations, carried out with the help of the QuTiP software package~\cite{Johansson2012, Johansson2013}, the Hilbert space is composed of 20 qubit and 120 resonator levels, with the qubit eigenstates calculated by prediagonalizing $\hat{H}_q$ in the 50-level harmonic-oscillator basis of the capacitive and inductive terms. The DASI algorithm step size is $\delta g/2\pi=1.5\,\textrm{MHz}$.

We start by fixing the external flux at $\Phi_\mathrm{ext}/\Phi_0=0.1$, where the $\ket{0}-\ket{4}$ transition frequency $\omega_{04}=(E_4 - E_0)/\hbar$ is close to being twice the bare resonator frequency as seen in Fig.~\ref{fig:levels}(a). This condition does not lead to features in the resonator dispersive shift but brings pairs of the states~$\ket{0,n}$ and $\ket{4,n-2}$ into a~resonance as depicted in Fig.~\ref{fig:levels}(b). Figure~\ref{fig:levels}(c) shows the evolution of the dressed eigenenergies from the corresponding bare states traced by the DASI algorithm for two particular dressed states, $\dket{0,40}$ and $\dket{4,38}$. We note the appearance of an~avoided level crossing, where the naïve diabatic labeling procedure with a single coarse step $\delta g$ equal to $g$ would intermix the state labels.

The dressed eigenenergies corresponding to qubit indices $k=0$ and $k=4$ and different photon numbers are shown in Fig.~\ref{fig:levels}(d) for $g/2\pi=150\,\mathrm{MHz}$. At small photon numbers, the spacing between the states corresponding to the qubit being in the ground ($k=0$) and the fourth excited state ($k=4$) is as large as $130\,\mathrm{MHz}$ while at photon numbers close to $n_\textrm{cross}=30$ this difference reduces to approximately $15\,\mathrm{MHz}$. This reduction in spacing, together with the level repulsion, indicates a~strong hybridization between states $\ket{0, n}$ and $\ket{4, n-2}$ at $n\sim n_{\rm cross}$, suggesting that a MIST could take place for a qubit prepared in its ground state at the resonator occupation $\bar{n} \sim n_{\rm cross}$.

Remarkably, the dispersive approximation for the ground state and parameters of Fig.~\ref{fig:levels} formally breaks down only at $n_{\rm crit, 0}\approx 90\approx 3n_{\rm cross}$, as calculated in Appendix~\ref{appendix:n_crit} using the generalization of $n_{\rm crit}$ to multilevel qubits~\cite{Dumas2024}. Therefore, the breakdown of the dispersive approximation is a poor indicator of potential MISTs as it can significantly overestimate the allowed value of $\bar{n}$. While $n_{\rm crit, 0}$ signifies strong hybridization between levels due to single-photon processes, the avoided-level-like crossing at $n_{\rm cross}$ happens due to multiphoton resonances, which are not captured by the second-order perturbation theory. For comparison, we also show the energies calculated in the dispersive approximation, which depend on $n$ linearly as $E_{\overline{k, n}} \approx E_{\overline{k, 0}} + n(E_{\overline{k, 1}} - E_{\overline{k, 0}})$ [thin gray lines in Fig.~\ref{fig:levels}(d)]. The energies agree reasonably well with the exact eigenenergies away from $n_{\rm cross}$ thanks to large $n_{\rm crit, 0}$. However, dispersive energies do not exhibit the anticrossing: at $n>n_{\rm cross}$, the level labels calculated using DASI and the dispersive approximation swap; see also Ref.~\cite{Dumas2024}.

 For the DASI step size $\delta g$ used throughout this paper, the algorithm may misidentify states at anticrossings with gaps $2g_{\rm eff}/2\pi \lesssim 1.6$ MHz; see Appendix~\ref{appendix:state_identification} for a more detailed discussion of the breakdown of DASI. In general, a step $\delta g$ can be chosen to accommodate a specific threshold $2g_{\rm eff}$ for anticrossings that need to be captured by the algorithm, making the algorithm very flexible. In particular, we find DASI to be more robust compared to the approach in which the states with one additional photon, i.e., $\dket{k, n+1}$, are identified by finding the maximum overlap with the states generated by $a^\dagger$ acting on already identified states $\dket{k,n}$~\cite{Shillito2022, Dumas2024}. The approach of Refs.~\cite{Shillito2022, Dumas2024} has been recently improved in the algorithm of Ref.~\cite{Goto2024}, where states $\dket{k, n+1}$ are labeled by comparing their bare-qubit occupations with those of states $\dket{k, n}$ and noticing that $E_{\overline{k, n+1}}$ must be close to $E_{\overline{k, n}} + \hbar \omega_r$. A disadvantage of DASI is its high computational cost, which, however, could be greatly reduced in the future by making the DASI step $\delta g$ change dynamically depending on the proximity to a level crossing.

\section{Metrics for identifying measurement-induced transitions} \label{Sec-metrics}

\begin{figure}
 \includegraphics[width=\columnwidth]{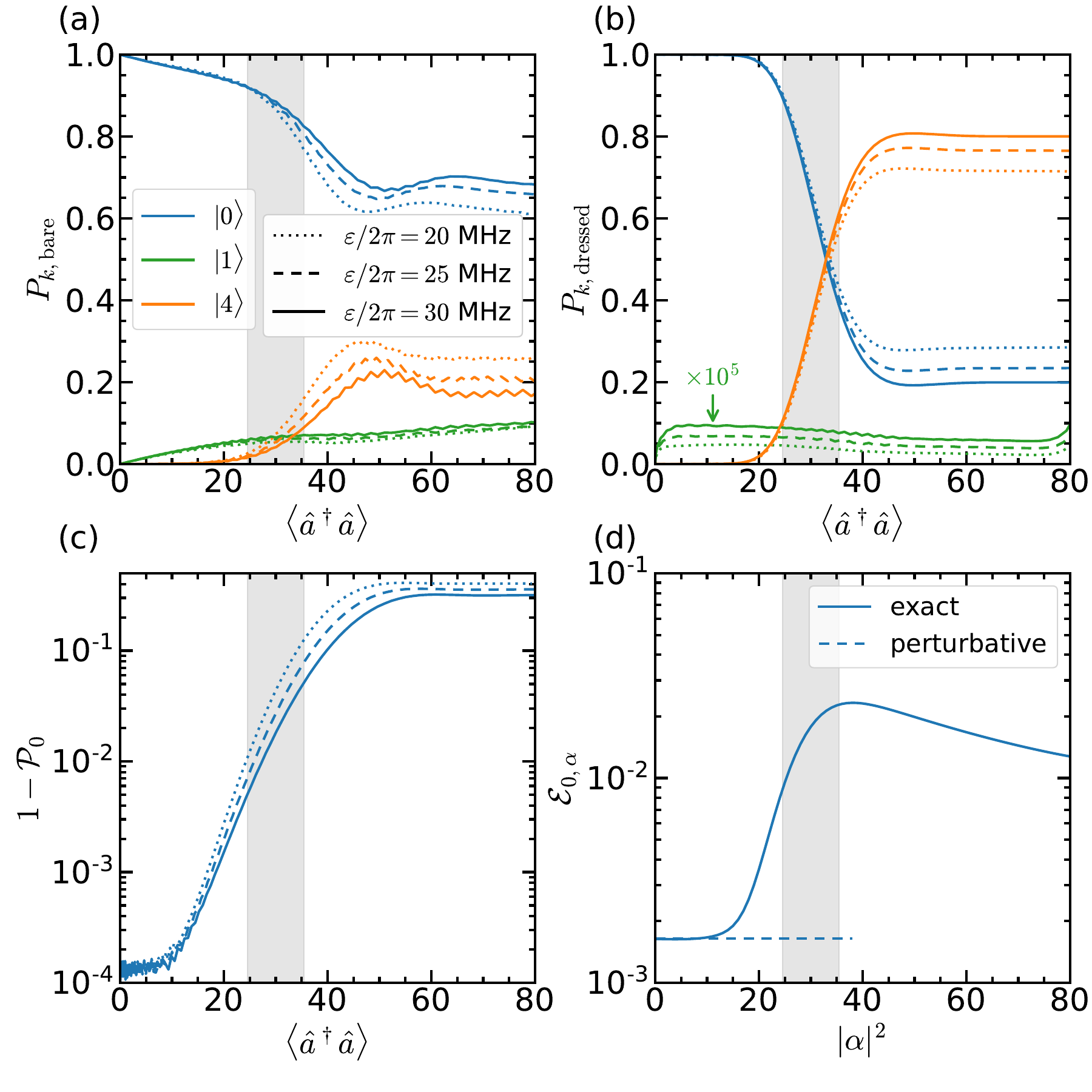}
 \caption{(a)--(c) Time evolution of the initial state $\dket{0, 0}$ in the presence of the resonant readout drive with $\varepsilon/2\pi=20$ (dotted lines), $25$ (dashed lines), and $30\,\textrm{MHz}$ (solid lines) for parameters of Fig.~\ref{fig:levels}.
 (a)~Probabilities $P_{0,\textrm{bare}}$ (blue), $P_{1,\textrm{bare}}$ (green), and $P_{4,\textrm{bare}}$ (orange) of finding the qubit in the bare states $\ket{0}$, $\ket{1}$, and $\ket{4}$ versus the average photon number $\expval{\hat{a}^\dagger \hat{a}}$. (b)~Same but for $P_{k,\textrm{dressed}}$, where $P_{k,\textrm{dressed}}$ is the total probability of finding the system in one of the states $\dket{k, n}$ (in the $k$-th dressed eigenladder).
 (c)~Qubit purity error $1-\mathcal{P}_0$ versus $\expval{\hat{a}^\dagger \hat{a}}$. (d)~Matrix-element error $\mathcal{E}_{0,\alpha}$ as defined by Eq.~\eqref{eq:error_mat_el} for the ground state versus the average photon occupation $|\alpha|^2$ calculated exactly (solid line) and perturbatively (dashed line). In all panels, gray shading highlights $n_\textrm{cross}\pm\sqrt{n_\textrm{cross}}$ with $n_{\rm cross}$ defined in Fig.~\ref{fig:levels}(d).}\label{fig:metrics}
\end{figure}

\subsection{Qubit occupation probabilities} \label{Sec-metrics-probabilities}

Time-domain simulations of Hamiltonian~\eqref{eq:model} with drive term~\eqref{eq:drive} reveal the expected MIST for parameters of Fig.~\ref{fig:levels}(d). 
We ignore any dissipation for simplicity and solve the Schr\"odinger equation for a drive with~$\omega_\textrm{drive}=\omega^{(0)}_r$ for the initial state $\dket{0, 0}$, where $\omega^{(k)}_r = (E_{\overline{k, 1}} - E_{\overline{k, 0}})/\hbar$ is the dressed resonator frequency for the qubit in $\ket{k}$.
To reduce the stray population of states $\dket{k, n}$ with a wrong qubit index $k\ne 0$, we choose an adiabatic drive~\cite{Khezri2016} whose strength increases as $\sin(\pi t/2t_{\rm raise})$ in the interval $0<t<t_{\rm raise}$ with $t_{\rm raise} = 20\,\textrm{ns}$ until it reaches a constant value~$\varepsilon$. In Figs.~\ref{fig:metrics}(a)--\ref{fig:metrics}(c), the simulation results are plotted versus the simultaneously computed average resonator photon occupation $\expval{\hat{a}^\dagger\hat{a}}$.

Figure~\ref{fig:metrics}(a) shows the probabilities of finding the qubit in various bare states defined as $P_{k,\textrm{bare}}\equiv\expval{\hat{\rho}_q}{k}=\sum_n\expval{\hat{\rho}}{k,n}$ for the $k$-th state, where $\hat{\rho}$ is the density matrix for the qubit-resonator system and 
\begin{equation}\label{rho_q}
\hat{\rho}_q = {\rm Tr}_{\rm res}\hat{\rho}
\end{equation}
is the reduced density matrix for the qubit with the trace taken over the resonator degree of freedom. We find that $P_{0,\textrm{bare}}$ noticeably decreases while $P_{1,\textrm{bare}}$ increases even with a few photons. This behavior is unrelated to MISTs since it is caused by the hybridization of bare states $\ket{0, n}$ and $\ket{1, n-1}$, which gets stronger with increasing~$n$. A~better metric is the probability of finding the interacting system in the $k$-th dressed eigenladder composed of dressed states $\dket{k, n}$ with a fixed $k$ (the dressed eigenladders are termed branches in Ref.~\cite{Dumas2024}). The physical interpretation of this is the probability of finding the qubit in its $k$-th dressed state. The dressed probabilities $P_{k,\textrm{dressed}}\equiv\sum_n\expval{\hat{\rho}}{\overline{k,n}}$ for $k=0,1$ and $4$ are shown in Fig.~\ref{fig:metrics}(b). We observe that $P_{0,\textrm{dressed}} \approx 1$ up to $\expval{\hat{a}^\dagger\hat{a}}\approx20$, while $P_{1,\textrm{dressed}}$ is suppressed by roughly five orders of magnitude, indicating that the system remains in the correct eigenladder. In contrast, $P_{0,\textrm{dressed}}$ drops quickly between $n_{\rm cross} - \sqrt{n_{\rm cross}}$ and $n_{\rm cross} + \sqrt{n_{\rm cross}}$ and flattens again beyond this region. The drop is accompanied by a sharp increase in $P_{4,\textrm{dressed}}$, indicating a population exchange between the dressed states with $k=0$ and $k=4$, a~clear sign of a MIST. Similar observations can be made for bare probabilities $P_{0,\textrm{bare}}$ and $P_{4,\textrm{bare}}$, although the effect is obscured and hard to quantify because of the initial hybridization-induced slope. 

A better indicator of MISTs is hidden in the dependence of bare probabilities on the drive power. We note that both $P_{k, \mathrm{bare}}$ and $P_{k, \mathrm{dressed}}$  are practically independent of $\varepsilon$ at $\expval{\hat{a}^\dagger\hat{a}} < 20$. They, however, depend on $\varepsilon$ at larger resonator occupation when the drive strength dictates how fast the system goes through the avoided-level-crossing-like region shaded in Fig.~\ref{fig:metrics}. A larger $\varepsilon$ reduces the adiabaticity of the evolution, i.e., the probability of traversing along the same eigenladder [the bottom eigenladder in Fig.~\ref{fig:levels}(d)], but increases the probability of a diabatic-like transition into the $k=4$ eigenladder [the top eigenladder in Fig.~\ref{fig:levels}(d)]. Therefore, the drive-strength order of the bare and dressed probabilities is reversed [compare the blue solid, dashed, and dotted lines in Figs.~\ref{fig:metrics}(a) and \ref{fig:metrics}(b)]. 

We stress that a MIST event during the resonator ring-up corresponds to the adiabatic passage through the avoided-level-like crossing in Fig.~\ref{fig:levels}(d). In this case,
the largest bare-state amplitude in the dressed state $\dket{0, n}$ evolves from $\ket{0, n}$ to $\ket{4, n-2}$, meaning that the qubit experiences a transition from its bare state $\ket{0}$ to bare state $\ket{4}$. Therefore, while $1 - P_{0, {\rm dressed}}$ can be thought of as the MIST probability before the shaded region in Fig.~\ref{fig:metrics}(b), it is $P_{0, {\rm dressed}}$ that corresponds to the MIST probability after that region, where $P_{0, {\rm dressed}}$ becomes flat, i.e., at $\expval{\hat{a}^\dagger\hat{a}} \gtrsim 40$. We note that the calculation of the experimentally relevant MIST probability requires simulation of the full readout process with the resonator ring-down and the dissipative dynamics, which is beyond the scope of this paper.

\subsection{Qubit purity}\label{Sec-metrics-purity}

Here, we discuss qubit-resonator entanglement by calculating the qubit purity $\mathcal{P}_k = {\rm Tr}\hat{\rho}_q^2$, where index $k$ stands for the initial state $\dket{k,0}$ at the start of a drive and $\hat{\rho}_q$ is given by Eq.~\eqref{rho_q}. Figure~\ref{fig:metrics}(c) shows the ``error'' $1-\mathcal{P}_0$ as a function of the average photon occupation $\expval{\hat{a}^\dagger\hat{a}}$. Due to qubit-resonator coupling, the purity of the initial state $\dket{0,0}$ is not exactly~$1$. Somewhat counterintuitively, $\mathcal{P}_0$ does not change appreciably during the initial stages of the drive despite the increased hybridized nature of the eigenstates $\dket{0, n}$ with $n\neq0$. In the MIST region, however, 
$1-\mathcal{P}_0$ grows rapidly from about $10^{-3}$ at $\expval{\hat{a}^\dagger\hat{a}}\approx20$ to about $10^{-1}$ at $\expval{\hat{a}^\dagger\hat{a}}\approx40$, indicating a stronger entanglement of the qubit-resonator system. Similar observations for the purity were made for the transmon~\cite{Shillito2022}. 

This behavior of the qubit-resonator entanglement is readily understood within the framework of the dressed coherent states~\cite{Sete2013, Sete2014, Govia2016}, which are defined for an arbitrary coherent-state amplitude $\alpha$ as
\begin{equation}
\dket{k, \alpha} = e^{-|\alpha|^2/2} \sum_n \frac{\alpha^n}{\sqrt{n!}}\dket{k, n}\label{eq:dressed}.
\end{equation}
A dressed coherent state approximates very well the state of an~interacting qubit-resonator system generated by a~classical microwave drive from the initial state $\dket{k, 0}$~\cite{Govia2016}. In comparison to the product state $\ket{k, \alpha} = \ket{k} \otimes \ket{\alpha}$, where $\ket{\alpha}$ is the resonator coherent state, Eq.~\eqref{eq:dressed}  shows that the system remains in the eigenladder for the $k$-th qubit state, in agreement with Fig.~\ref{fig:metrics}(b). As shown for the transmon, the states evolve primarily within the qubit eigenladder even when $|\alpha|^2> n_{\rm crit}$ with two corrections to Eq.~\eqref{eq:dressed}: small leakage to neighboring eigenladders and squeezing of the correct eigenladder portion, caused by the qubit nonlinearity~\cite{Khezri2016}. Importantly, in stark contrast with strongly hybridized eigenstates $\dket{k, n}$, dressed coherent states~\eqref{eq:dressed} remain practically unentangled: for a large $|\alpha|^2$, $\dket{k, \alpha} \approx \ket{q}\otimes \ket{\alpha}$ with some $\ket{q}=\sum_k c_k \ket{k}$~\cite{Khezri2016}. This also holds when Eq.\eqref{eq:dressed} is corrected for squeezing~\cite{Khezri2016}.
Thus, purity is expected to remain close to unity in the dispersive and strong-hybridization limits, provided the system predominantly stays in the correct eigenladder. In Appendix~\ref{appendix:purity}, we illustrate this statement for Jaynes-Cummings Hamiltonian by calculating $\mathcal{P}_0$ analytically. In particular, we show that in the dispersive limit, the deviation of $\mathcal{P}_0$ from 1 is the second-order effect for eigenstates, but is the effect of only \emph{sixth} order in the coupling strength for dressed coherent states defined by Eq.~\eqref{eq:dressed}. 

The initially weak increase in purity error $1-\mathcal{P}_0$ seen in Fig.~\ref{fig:metrics}(c) is consistent with the expectation of the qubit-resonator system to remain in a state close to $\dket{0, \alpha}$ and thus be almost unentangled. The subsequent rapid growth of $1-\mathcal{P}_0$ by two orders of magnitude in the MIST region is caused by a probability splitting between the $k=0$ and $k=4$ eigenladders [see Fig.~\ref{fig:metrics}(b)], which  breaks the dressed-coherent-state picture. The onset of this transition can be determined by computing the sensitivity of the purity error to the drive amplitude at a~fixed $\expval{\hat{a}^\dagger \hat{a}}$, although the non-RWA terms can complicate the numerical evaluation of these sensitivities. We note that in the presence of dissipation, $1-\mathcal{P}_0$ behaves similarly to Fig.~\ref{fig:metrics}(c) with the effect of an increased resonator decay rate being analogous to that of a smaller drive power, which reduces the speed of traversing through the avoided-level-crossing-like region; see Appendix~\ref{appendix:dissipation}.

Although both the probability of finding the system in a~particular eigenladder [Fig.~\ref{fig:metrics}(b)] and the qubit purity [Fig.~\ref{fig:metrics}(c)] are sensitive MIST metrics, calculation of $P_{k, \mathrm{dressed}}$ is computationally more intensive as it requires both time-domain simulations and identification of dressed states for large photon numbers with a defined labeling algorithm. In comparison, $\mathcal{P}_0$ can be computed directly from the time-dependent density matrix $\hat{\rho}(t)$, expressed in the bare basis, eliminating the need to label dressed eigenstates. Therefore, the qubit purity (and other entanglement measures) can be used as a practical probe of MISTs when doing the time domain simulations. We note that a probability metric based on the occupation of approximate eigenstates in the dispersive approximation, corresponding to the gray lines of Fig.~\ref{fig:levels}(d), may also be practical in certain situations but can fail when $n_{\rm cross} > n_{{\rm crit}, k}$.

\subsection{Dressed matrix elements of the drive}

The exact shape of the purity-error curves depends on the resonator ring-up protocol and photon loss. Here, we introduce another metric based on the matrix elements of $\hat{a}$ that does not require time-domain simulations with a specific drive term. Instead, we now assume that the qubit-resonator system is already in a~dressed coherent state $\dket{k, \alpha}$ and check whether the dressed-coherent-state picture remains valid if an extra photon is added or removed. To this end, we define a~simple error metric that quantifies to what extent the dressed coherent state is an eigenstate of bare $\hat{a}$:
\begin{equation}\label{eq:error_mat_el}
    \mathcal{E}_{k, \alpha} = \left| 1 - \frac{\expval{\hat{a}}{\overline{k,\alpha}}}{\alpha}\right|\,.
\end{equation}
The metric is trivially zero for bare coherent states $\ket{k,\alpha}$ since $\hat{a}\ket{k,\alpha} = \alpha \ket{k,\alpha}$. For dressed states, it shows how close the operators $\hat{a}^\dagger$ and $\hat{a}$ are to their dressed versions $\hat{\bar{a}}^\dagger$ and $\hat{\bar{a}}$, where $\hat{\bar{a}} = \sum_{k, n}\sqrt{n+1}\dket{k, n} \dbra{k, n+1}$. It is the condition $\hat{a} \approx \hat{\bar{a}}$ that results in the generation of dressed coherent states by $\hat{V}_{\rm drive}$~\cite{Khezri2016}; therefore, a large $\mathcal{E}_{k, \alpha}$ indicates the breakdown of the dressed-coherent-state picture and thus the onset of MISTs.

We note that to compute $\mathcal{E}_{k, \alpha}$ numerically, special attention has to be paid to the signs of eigenstates at each step of the DASI algorithm: in addition to identifying eigenstates $\dket{k, n}_{g+\delta g}$ by maximizing overlaps, we also fix their phases by requiring $\sideset{_g}{_{g+\delta g}}{\mathop{\dmel{k, n}{k, n}}}$ to have positive real values. In this way, the numerically calculated $\dket{k, \alpha}$ is well defined and is not spoiled by the potential phase ambiguity of $\dket{k, n}$.

In Fig.~\ref{fig:metrics}(d), we plot $\mathcal{E}_{0,\alpha}$ as a function of $|\alpha|^2$ computed for the same parameters as in other panels (solid line). We find an order-of-magnitude jump between $|\alpha|^2\approx20$ and $|\alpha|^2\approx40$, indicating the expected MIST in agreement with the time-domain simulations. Due to state hybridization,  $\mathcal{E}_{k, 0}$ is not zero at $|\alpha|^2=0$ with the actual value closely matching the perturbative result, which is independent of $|\alpha|$ [dashed line in Fig.~\ref{fig:metrics}(d); calculated in Appendix~\ref{appendix:perturbation_mat_el}].

The practical advantage of the metric defined by Eq.~\eqref{eq:error_mat_el} is that it requires only time-independent terms of the Hamiltonian~\eqref{eq:model} and allows direct comparison of different qubit types regardless of the specifics of the drive and system losses. 

\section{Dependence on Hamiltonian parameters}\label{Sec-dependence-on-parameters}

\begin{figure}
 \includegraphics[width=\columnwidth]{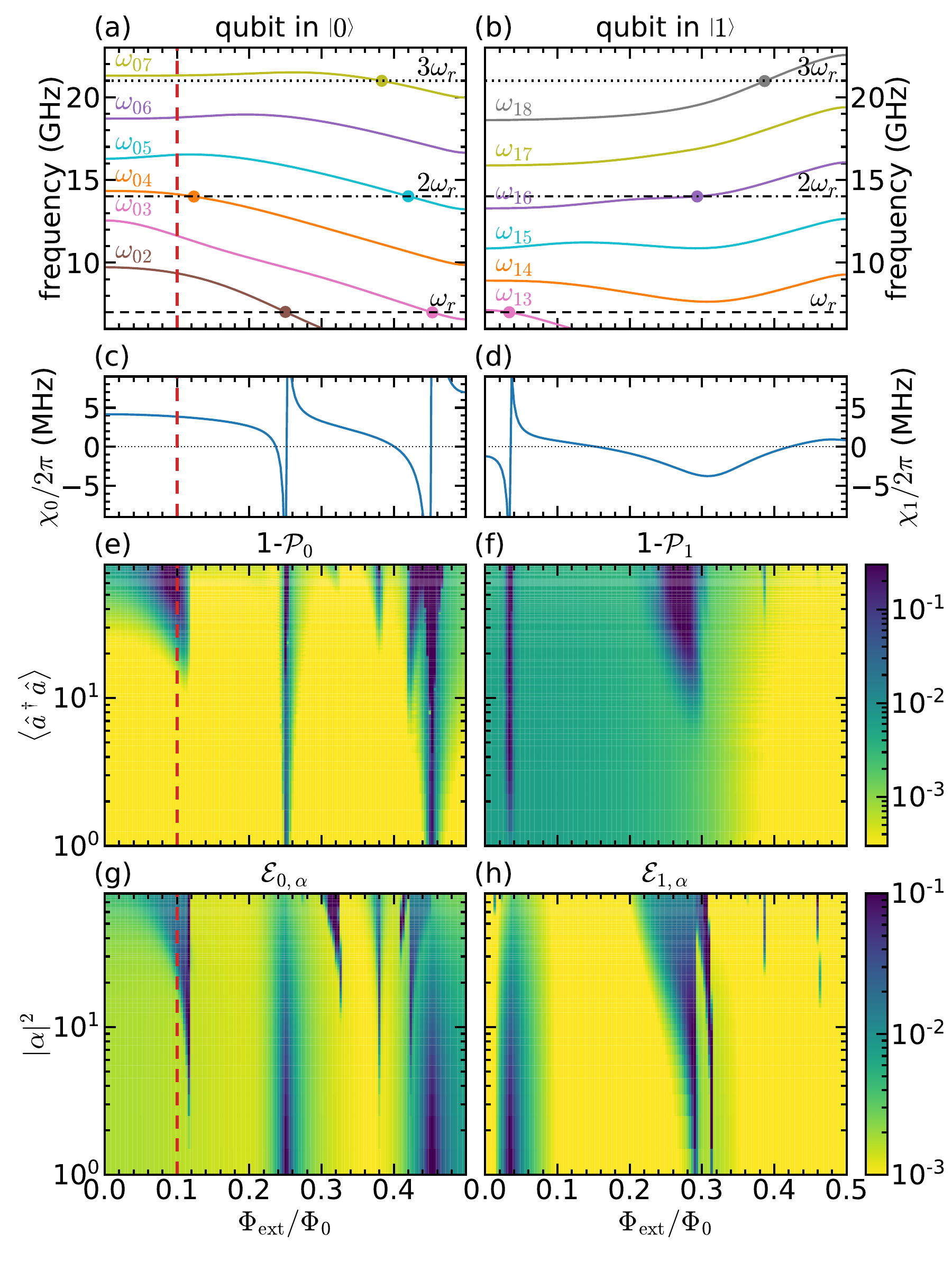}
 \caption{MIST dependence on the external flux~$\Phi_\mathrm{ext}$ (horizontal axes) for a fluxonium initially in the ground (left) and first excited (right) states. (a),~(b) Bare qubit and resonator transition frequencies with the resonance conditions highlighted by round markers.
 (c),~(d) Dispersive shifts $\chi_k = (E_{\overline{k, 1}} - E_{\overline{k, 0}})/\hbar - \omega_r$, which diverge when $\omega_r$ matches one of the qubit transitions~$\omega_{kk'}$. (e),~(f) Qubit purity errors $1-\mathcal{P}_k$ versus the average photon occupation $\expval{\hat{a}^\dagger\hat{a}}$ (vertical axes). (g),~(h) Matrix-element errors~$\mathcal{E}_{k,\alpha}$ versus the average photon occupation $|\alpha|^2$ (vertical axes). The flux value of Figs.~\ref{fig:levels} and \ref{fig:metrics} is highlighted by vertical dashed lines in the left-column panels; other parameters are the same as in Figs.~\ref{fig:levels} and \ref{fig:metrics}.}\label{fig:phi_ext}
\end{figure}

\begin{figure}
    \centering
    \includegraphics[width=\columnwidth]{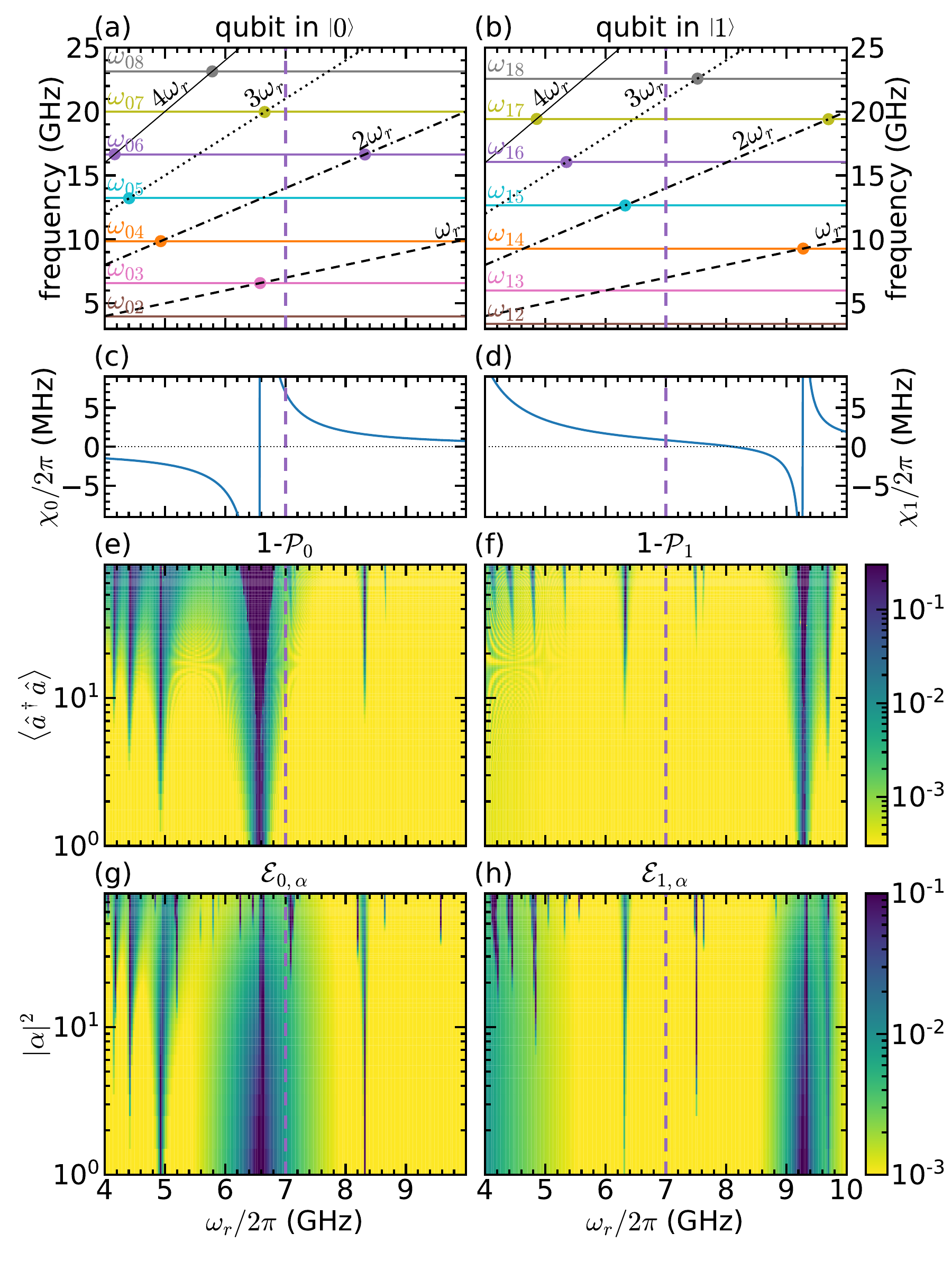}
    \caption{Same as Fig.~\ref{fig:phi_ext}, but for the dependence on the resonator frequency $\omega_r$ (horizontal axes) at the $\Phi_{\rm ext} = \Phi_0/2$ sweet spot. The value $\omega_r/2\pi = 7.0\,\textrm{GHz}$ of Fig.~\ref{fig:phi_ext} is highlighted by vertical dashed lines (the cross sections along these lines are identical to the rightmost cross sections in the corresponding panels of Fig.~\ref{fig:phi_ext}). Round markers in panels~(a) and (b) label only those resonances for which the sweet-spot selection rules allow state transitions.}
    \label{fig:omega_r}
\end{figure}

Here, we apply the ideas from Sec.~\ref{Sec-metrics} to explore MISTs in a wider parameter space of the chosen fluxonium circuit. In Fig.~\ref{fig:phi_ext}, we investigate the dependence on the external flux~$\Phi_\textrm{ext}$. We show the simulation results as a~function of $\Phi_{\rm ext}$ for a qubit prepared in states $\ket{0}$ (left column) and $\ket{1}$ (right column). To build an intuition where MISTs are possible, we start by plotting single-qubit transition frequencies $\omega_{kk'} = (E_{k'} - E_{k})/\hbar$ for $k=0$ and $k=1$ in Figs.~\ref{fig:phi_ext}(a) and \ref{fig:phi_ext}(b). In these figures, the round markers indicate where $\omega_{kk'}$ cross multiples of the bare resonator frequency $\omega_r$. ``First-order'' crossings $\omega_{kk'} = \omega_r$ lead to divergences in the corresponding dispersive shift $\chi_k = \omega^{(k)}_r - \omega_r$ of the resonator frequency~\cite{Zhu2013a}, clearly observed in Figs.~\ref{fig:phi_ext}(c) and \ref{fig:phi_ext}(d). ``Higher-order'' crossings, e.g., those with $2\omega_r$ and $3\omega_r$, do not cause the dispersive shift to diverge but are expected to cause MISTs. Ultimately, the anticrossing of the $E_{\overline{0, n}}$ and $E_{\overline{4, n-2}}$ ladders in Fig.~\ref{fig:levels}(d) occurs at a relatively small $n_{\rm cross}=30$ because $\omega_{04}$ is close to $2\omega_r$ [see Fig.~\ref{fig:levels}(a) and the orange circle in Fig.~\ref{fig:phi_ext}(a)]. 

The intuition built upon the crossings of bare qubit and resonator frequencies is confirmed by simulations of the qubit purity~$\mathcal{P}_k$ and the matrix-element error~$\mathcal{E}_{k, \alpha}$. In Figs.~\ref{fig:phi_ext}(e) and \ref{fig:phi_ext}(f), we show the qubit purity errors $1-\mathcal{P}_0$ and $1-\mathcal{P}_1$ for the starting states $\dket{00}$ and $\dket{10}$ as two-dimensional color maps versus $\Phi_{\rm ext}$ and $\expval{\hat{a}^\dagger \hat{a}}$. Both errors are simulated for $\varepsilon/2\pi = 25$ MHz as described in Sec.~\ref{Sec-metrics} except that the drive frequency is $\omega_{\rm drive} = \omega_r^{(k)}$ for the initial state $\dket{k0}$ to simplify simulations (the time-domain simulations do not correspond to an actual readout where $\omega_{\rm drive}$ is not qubit-state dependent). In Figs.~\ref{fig:phi_ext}(g) and \ref{fig:phi_ext}(h), we show $\mathcal{E}_{0, \alpha}$ and $\mathcal{E}_{1, \alpha}$ versus $\Phi_{\rm ext}$ and $|\alpha|^2$. We brief on several observations. First, the features (i.e., the regions with increased errors) in the $1-\mathcal{P}_k$ panels mostly agree with the features in the $\mathcal{E}_{k, \alpha}$ panels. Second, the largest features in these panels, which occur at small photon numbers $\expval{\hat{a}^\dagger \hat{a}}$ and $|\alpha|^2$, correspond to divergences in the dispersive shifts. Third, the features appearing at crossings of single-qubit frequencies with $2\omega_r$, i.e., at $\Phi_{\rm ext}/\Phi_0 \approx 0.1$ for a qubit in $\ket{0}$ and at $\Phi_{\rm ext}/\Phi_0 \approx 0.3$ for a qubit in $\ket{1}$, are more pronounced with an earlier onset of MISTs in comparison to the features corresponding to crossings with $3\omega_r$. 

Figure~\ref{fig:omega_r} shows the same metrics calculated for the same parameters as Fig.~\ref{fig:phi_ext} but as a function of the resonator frequency~$\omega_r$ for the fluxonium parked at its half-integer sweet spot, $\Phi_{\rm ext}/\Phi_0=0.5$. We note that due to the selection rules at the sweet spot, i.e., $\bra{k}\hat{n}\ket{k'}=0$ for two states $k$ and $k'$ of the same parity, not every frequency collision observed in Figs.~\ref{fig:omega_r}(a) and~\ref{fig:omega_r}(b) leads to a divergence in the dispersive shift or an error growth in Figs.~\ref{fig:omega_r}(e)--\ref{fig:omega_r}(h). Even at higher orders, $\hat{V}_{qr}(g)$ mixes only the bare states $\ket{k,n}$ of the same combined parity $(-1)^{k+n}$ and, therefore, the parity of the dressed states is well defined. The avoided-level-like crossings such as the one shown in Fig.~\ref{fig:levels}(d) occur only between dressed states of the same parity, and, thus, the resonance condition $\omega_{kk'}=n\omega_r$ must be supplemented by the parity requirement $(-1)^{k-k'}=(-1)^{n}$. Only such crossings are highlighted by round markers in Figs.~\ref{fig:omega_r}(a) and~\ref{fig:omega_r}(b), and the onset of the errors for each of these crossings is clearly visible in Figs.~\ref{fig:omega_r}(e)--\ref{fig:omega_r}(h).

We emphasize that except for the locations of divergences in the dispersive shifts, features in the heat maps of Figs.~\ref{fig:phi_ext} and \ref{fig:omega_r} are not related to the formal breakdown of the dispersive approximation; see Appendix~\ref{appendix:ncrit_vs_parameters} for the calculations of $n_{{\rm crit}, k}$ versus $\Phi_{\rm ext}$ and $\omega_r$. Finally, we stress that the two metrics are not fully equivalent and care should be taken when comparing the photon numbers at which a particular feature becomes pronounced. While $\mathcal{E}_{k, \alpha}$ is agnostic to a specific ring-up protocol, $1-\mathcal{P}_k$ is a characteristic of both the spectrum and the readout protocol and should exhibit the same onset of features as $\mathcal{E}_{k, \alpha}$ only for an adiabatic traversal through the avoided-level-like crossings. A finite drive strength~$\varepsilon$ used in the time-domain simulations can result in diabatic passages through the avoided-level-like crossings, shifting the appearance of features in $1-\mathcal{P}_k$ to higher photon numbers compared to $\mathcal{E}_{k,\alpha}$. The sensitivity of $1-\mathcal{P}_k$ to $\varepsilon$ and the connection between various MIST metrics are  discussed in more detail in Appendix~\ref{appendix:comparison}.

\section{Summary and conclusions}\label{Sec-conclusion}

Despite the well-established techniques for engineering dispersive shifts and coupling losses in circuit QED architectures, the task of readout optimization remains partially a trial-and-error empirical endeavor. Crucially for this task, it is possible for qubit-resonator systems to have the same dispersive shifts and loss rates but behave differently with the increasing number of readout photons because of the measurement-induced state transitions. In this paper, we demonstrated how the qubit purity and the matrix-element error computed for the dressed coherent states can be employed to identify parameter regimes favorable for implementing fast and high-fidelity readout protocols compatible with the deterministic reset. 

The first metric, the qubit purity, requires the time-domain simulations of the readout process to quantify the entanglement, while the second metric, the matrix-element error, employs accurate identification of dressed states and quantifies the ease with which the drive can increase the size of the dressed coherent state. Even though the system losses and the drive specifications do not enter the definition of the matrix-element error, we expect this metric to be useful for identifying high-power readout regimes that are robust to minor drive calibration errors. Both metrics do not rely on any approximation, such as the RWA, apart from the presumed (and practically desired) validity of the dressed-coherent-state picture. The metrics can be extended to the dressed squeezed states and evaluated along the semiclassical trajectories the drive may produce in the resonator phase space. They allow a quantifiable comparison of different superconducting qubit types with respect to their performance in dispersive-readout schemes and can provide universal guidelines for readout optimization. 

\section*{Acknowledgments}
K.N. thanks the Atlantic Quantum team, especially Youngkyu Sung and Bharath Kannan, for useful discussions. This research was supported in part through computational resources provided by Syracuse University.
Dissipative dynamics computations in Appendix~\ref{appendix:dissipation} were performed on the Syracuse University OrangeGrid supported by the National Science Foundation Grant No. ACI-1341006. This material is based upon work supported by the National Science Foundation under Grant No. 2412597.

\appendix

\section{Breakdown of the dispersive approximation}
\label{appendix:dispersive}

\subsection{Critical photon number}
\label{appendix:n_crit}

\begin{figure}
    \centering
    \includegraphics[width=\columnwidth]{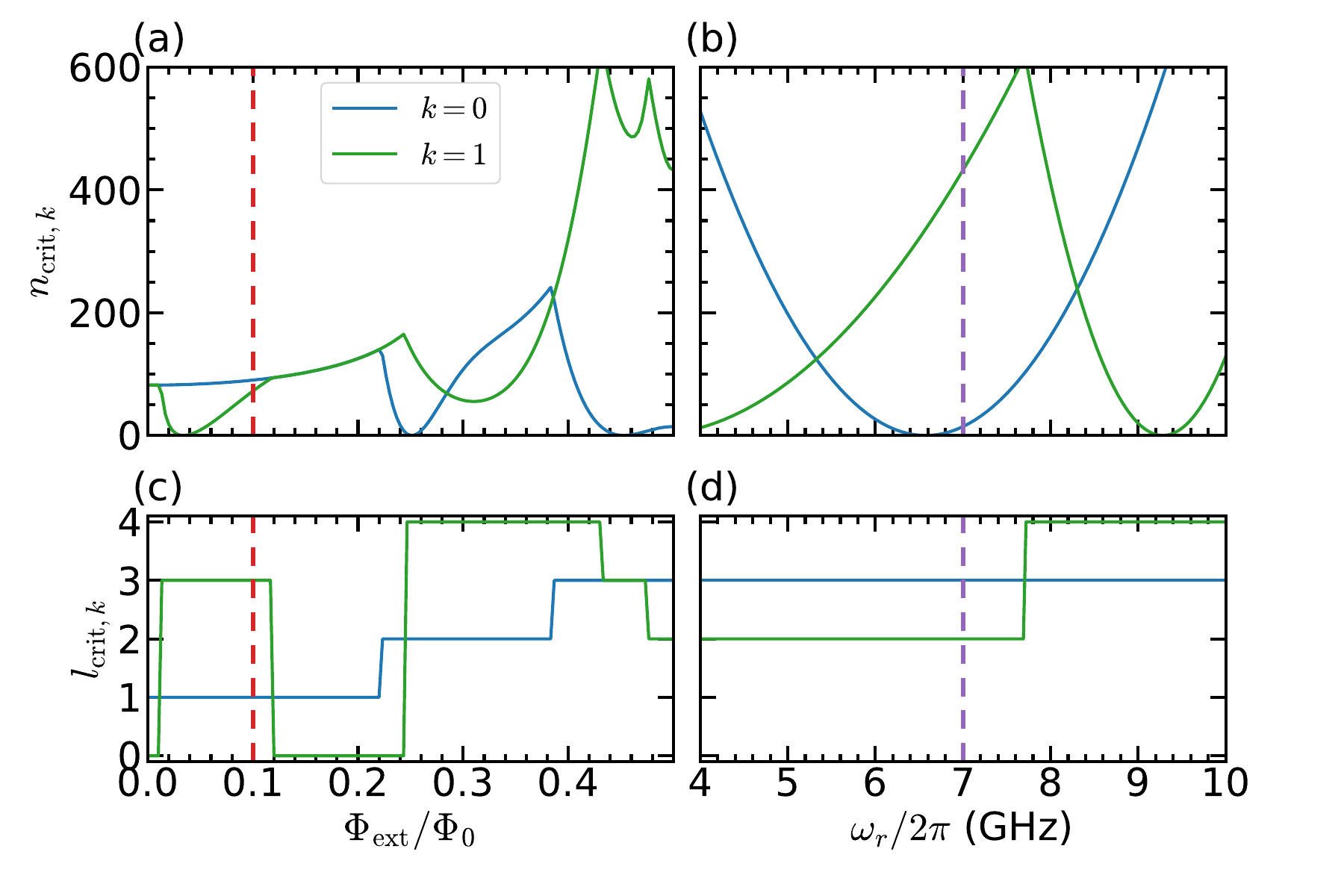}
    \caption{(a), (b) Dispersive critical photon numbers $n_{\mathrm{crit}, k}$ and (c), (d)~qubit state indices $l_{\mathrm{crit}, k}$ for the transitions $\ket{k}\to \ket{l_{\mathrm{crit}, k}}$ defining $n_{\mathrm{crit}, k}$ for a qubit in its ground ($k=0$, blue lines) and first excited ($k=1$, green lines) states. The left column shows the dependence on $\Phi_\mathrm{ext}$ for parameters of Fig.~\ref{fig:phi_ext}, while the right column shows the dependence on $\omega_r$ for parameters of Fig.~\ref{fig:omega_r}. Vertical dashed lines mark the same values as they do in Figs.~\ref{fig:phi_ext} and \ref{fig:omega_r}.
    } 
    \label{fig:ncrit}
\end{figure}

We define the critical photon number corresponding to the breakdown of the conventional dispersive approximation using the generalization of $n_{\rm crit} = \Delta^2/4g^2$ to multilevel qubits~\cite{Dumas2024}. For a qubit in state $\ket{k}$, $n_{\rm crit}$ is given by
\begin{equation}\label{ncrit}
    n_{{\rm crit}, k} = \min_{l\ne k}\left|\frac{|\omega_{kl}| - \omega_r}{2g\mel{k}{\hat{n}}{l}}\right|^2,
\end{equation}
where the minimum is taken only over those qubit levels $l\ne k$ for which $\mel{k}{\hat{n}}{l} \ne 0$. 

For the parameters of Fig.~\ref{fig:levels}, Eq.~\eqref{ncrit} gives $n_{{\rm crit}, 0}\approx90$, which is much larger than the avoided-level-crossing location at $n_{\rm cross}=30$. There is no contradiction here: $n_{{\rm crit}, k}$ signifies the resonator photon occupation~$n$ when the first-order correction to the qubit-resonator state $\ket{k, n}$ and the second-order correction to energy $E_{k, n}$ become large, while the avoided-level-like crossing in Fig.~\ref{fig:levels}(d) can be traced to the breakdown of the second-order correction to $\ket{0, n}$ and fourth-order correction to $E_{0, n}$. 
The opposite situation, when $n_{\rm cross} > n_{{\rm crit}, k}$, is also possible~\cite{Sank2016}, demonstrating that $n_{{\rm crit}, k}$ is generally a bad predictor of the breakdown of the higher-order corrections and MISTs. We note that while the location of the anticrossing can be associated with a divergent term in perturbation theory, the anticrossing of the levels itself is a nonperturbative effect~\cite{Dumas2024}.

\subsection{Dependence on Hamiltonian parameters}
\label{appendix:ncrit_vs_parameters}

Figure~\ref{fig:ncrit} shows $n_{{\rm crit}, 0}$ and $n_{{\rm crit}, 1}$ for the same parameters as in Figs.~\ref{fig:phi_ext} and~\ref{fig:omega_r} and the values of $l_{{\rm crit}, 0}$ and $l_{\mathrm{crit}, 1}$ that correspond to the qubit indices $l$ that minimize Eq.~\eqref{ncrit} for $k=0$ and $k=1$. As expected, the values of $n_{{\rm crit}, k}$ correlate strongly with the dispersive shifts $\chi_k$ shown in Figs.~\ref{fig:phi_ext} and \ref{fig:omega_r}: $n_{{\rm crit}, k}$ vanishes whenever $\chi_k$ diverges. Equation~\eqref{ncrit} makes it evident that $n_{{\rm crit}, k}=0$ at the single-photon resonance conditions $|\omega_{kl}| = \omega_r$, when the second-order corrections to energies $E_{k, n}$ diverge at any $n\ge 0$ (for $l<k$) or $n \ge 1$ (for $l>k)$, resulting in large shifts $\chi_k$. In comparison, there is no visible correlation between $n_{{\rm crit}, k}$ and MIST-related features in Figs.~\ref{fig:phi_ext} and \ref{fig:omega_r}, which are due to multiphoton resonances.

\section{Breakdown of the state-identification algorithm}
\label{appendix:state_identification}

\begin{figure}
    \centering
    \includegraphics[width=\columnwidth]{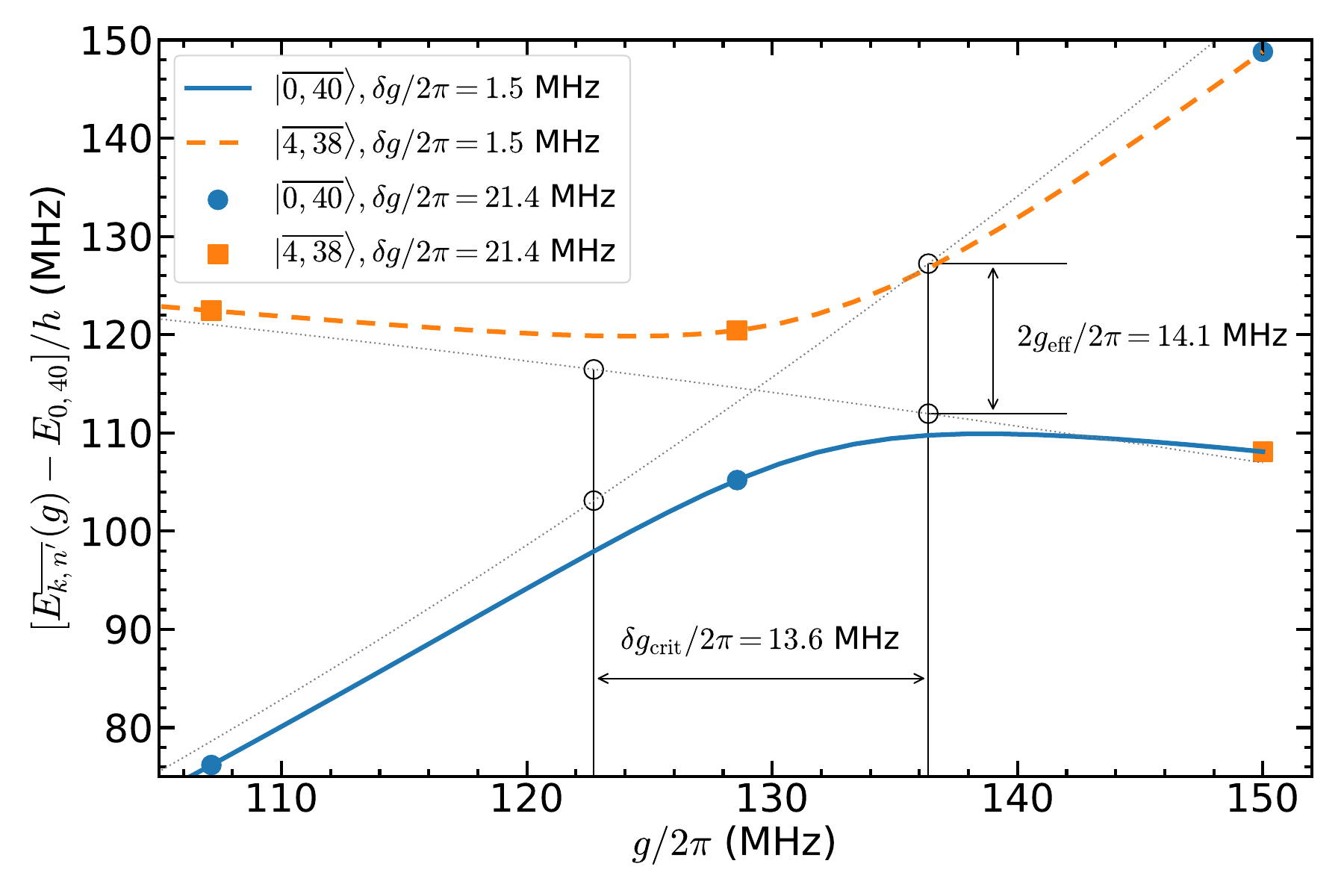}
    \caption{Zoom-in into Fig.~\ref{fig:levels}(c) near the avoided level crossing. Blue circles and orange squares show energies of the same states as solid and dashed lines but labeled using a larger DASI step $\delta g/2\pi = 21.4\,\mathrm{MHz}$. The critical DASI step $\delta g_{\rm crit}$ is estimated using the energies calculated in the dispersive approximation (gray dots) and the gap $2g_{\rm eff}$ at the avoided level crossing; see Eq.~\eqref{eq:dasi-breakdown}.} 
    \label{fig:dasi}
\end{figure}

The DASI algorithm presented in Sec.~\ref{sec-model-dasi} fails when the energy gap $2g_{\rm eff}$ at the anticrossing such as that shown in Fig.~\ref{fig:levels}(c) is too small or the step size $\delta g$ used by the algorithm is too large. We illustrate this breakdown in Fig.~\ref{fig:dasi}, where we zoom in on the avoided level crossing of Fig.~\ref{fig:levels}(c) to show the energies of the states identified using a large $\delta g/2\pi = g/7 \approx 21.4\,\mathrm{MHz}$; see the blue circles and orange squares. To better understand the DASI failure, we consider a simplified model of a two-level system described by
\begin{equation}\label{eq:model-Landau-Zener}
    \hat{H} = \begin{pmatrix}
    -\Delta_{\rm diab}(g)/2 & g_{\rm eff}(g) \\
    g_{\rm eff}(g) & \Delta_{\rm diab}(g)/2
    \end{pmatrix}\,.
\end{equation}
Here, using the terminology of the Landau-Zener problem, $\Delta_{\rm diab}$ and $g_{\rm eff}$ are the $g$-dependent energy difference and interaction between diabatic states $\ket{-}$ and $\ket{+}$. The eigenstates of this Hamiltonian are given by
\begin{equation}
\begin{aligned}
        \dket{-}_g &= \cos\theta_g \ket{-} - \sin\theta_g \ket{+}\,,\\
        \dket{+}_g &= \sin\theta_g \ket{-} + \cos\theta_g \ket{+}\,,
\end{aligned}
\end{equation}
where $\theta_g = \frac{1}{2}\tan^{-1}[2g_{\rm eff}(g)/\Delta_{\rm diab}(g)]$ at $\Delta_{\rm diab}(g) > 0$ (before the anticrossing) and $\theta_g = \pi/2 - \frac{1}{2}\tan^{-1}[2g_{\rm eff}(g)/|\Delta_{\rm diab}(g)|]$ at $\Delta_{\rm diab}(g) < 0$ (after the anticrossing). The overlaps between these eigenstates at one step of the DASI algorithm are given by
\begin{equation}
    \begin{aligned}
        \sideset{_g}{_{g+\delta g}}{\mathop{\dmel{-}{-}}} &= \cos(\theta_{g+\delta g} - \theta_{g})\,,\\
        \sideset{_g}{_{g+\delta g}}{\mathop{\dmel{-}{+}}} &= \sin (\theta_{g+\delta g} - \theta_{g})\,,
    \end{aligned}
\end{equation}
implying that the algorithm misidentifies states when $|\tan(\theta_{g+\delta g} - \theta_{g})| \ge 1$. Assuming, for simplicity, that $g$ and $g+\delta g$ are chosen symmetrically around the avoided level crossing, so $g_{\rm eff}(g)/|\Delta_{\rm diab}(g)| = g_{\rm eff}(g+\delta g)/|\Delta_{\rm diab}(g+\delta g)|$ and $\pi/4 - \theta_g = \theta_{g+\delta g} - \pi/4$, we find the breakdown condition $2\theta_g \le \pi/4$, or 
\begin{equation}\label{eq:dasi-breakdown}
    \Delta_{\rm diab}(g) \ge 2g_{\rm eff}(g)\,.
\end{equation}
Following this criterion, we estimate the critical DASI step $\delta g_{\rm crit}$ for parameters of Fig.~\ref{fig:dasi} by using the energies obtained in the dispersive approximation to find $\Delta_{\rm diab}(g)$ (gray dotted lines) and by extracting $2g_{\rm eff}/2\pi \approx 14.1\,\mathrm{MHz}$ from the actual gap at the avoided level crossing, where we ignore the change in $g_{\rm eff}(g)$ in the interval of interest. The estimate yields $\delta g_{\rm crit}/2\pi \approx 13.6\,\mathrm{MHz}$, as illustrated in the figure. Alternatively, for the $\delta g/2\pi = 1.5\,\mathrm{MHz}$ step of the main text, we estimate that the algorithm may break down at $2g_{\rm eff}/2\pi \sim 1.6\,\mathrm{MHz}$.

\section{Qubit purity for two-level qubits}
\label{appendix:purity}

\subsection{Model}

\begin{figure}
    \centering
    \includegraphics[width=\columnwidth]{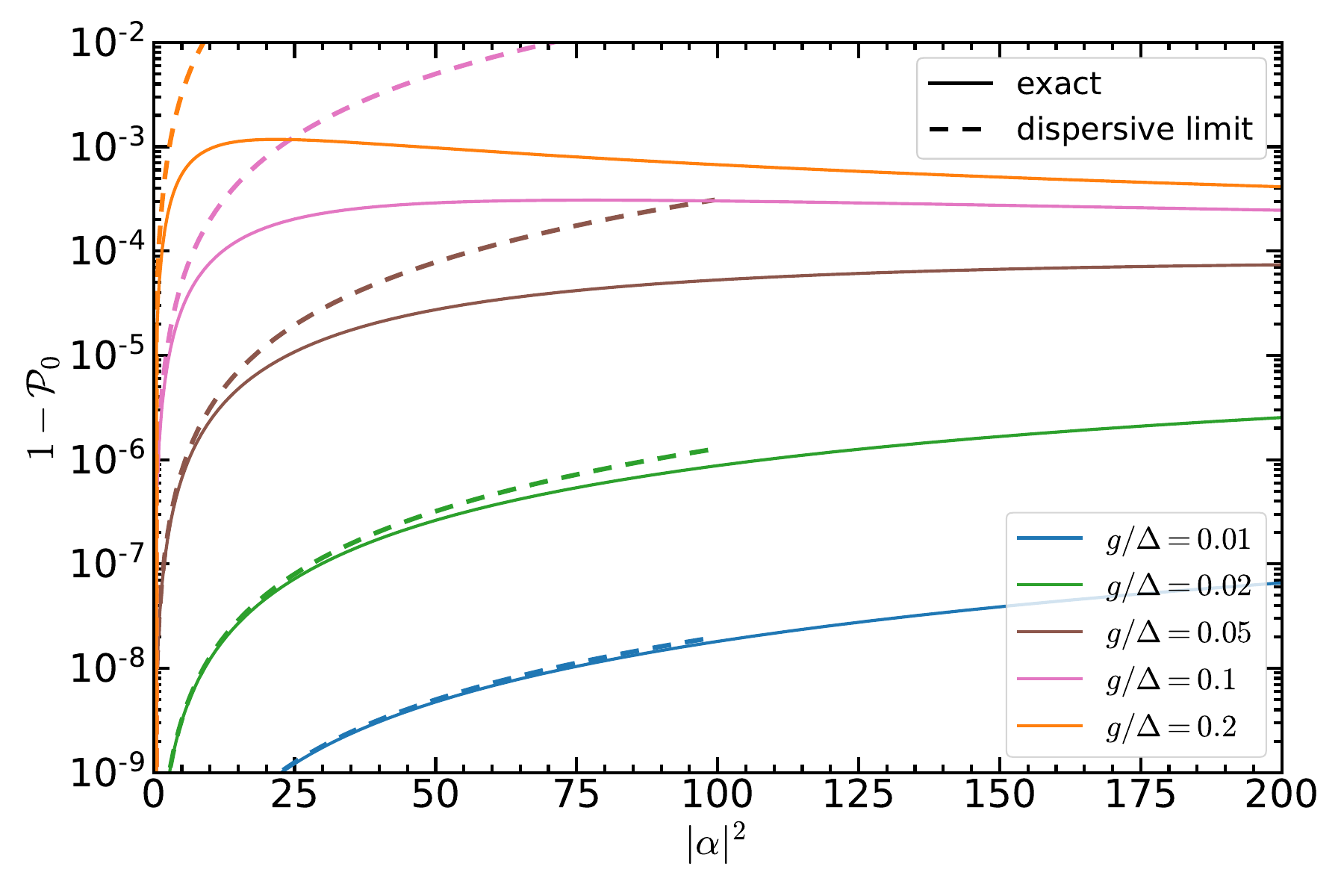}
    \caption{Purity error $1-\mathcal{P}_0$ calculated for the dressed coherent states $\dket{0, \alpha}$ in the Jaynes-Cummings model~\eqref{model-two-level} versus $|\alpha|^2$. The qubit purity is calculated exactly (solid lines) and using the dispersive result of Eq.~\eqref{purity_0alpha_dispersive} (dashed lines) for various values of the dispersive parameter $g/\Delta$.}
    \label{fig:purity}
\end{figure}

Here, we build intuition for why the purity of the reduced qubit density matrix remains close to 1 for dressed coherent states~\eqref{eq:dressed} by considering a simple Jaynes-Cummings model in the RWA:
\begin{equation}\label{model-two-level}
    \frac{\hat{H}_{\rm JC}}{\hbar} =   \omega_r \hat{a}^\dagger \hat{a} - \frac{\omega_q}{2}\hat{\sigma}_z + g\left(\hat{a}\hat{\sigma}_+ + \hat{a}^\dagger\hat{\sigma}_-\right)\,.
\end{equation}
$\omega_q$ and $\omega_r$ are the bare qubit and resonator frequencies, and $g$ is the coupling strength. The purity error $1-\mathcal{P}_0$ calculated numerically for this model is shown in Fig.~\ref{fig:purity} for various values of dimensionless coupling strength $\lambda = g/\Delta$, where $\Delta = \omega_q - \omega_r$. In this section, we calculate $\mathcal{P}_0$ for dressed coherent states analytically and derive a~simple expression in the dispersive limit.

The Hamiltonian~\eqref{model-two-level} has a~block-diagonal structure, with each block corresponding to a different RWA strip defined by the total excitation number. The exact diagonalization of the $2\times 2$ block with $n>0$ excitations gives the relation between dressed and bare states:
\begin{equation}\label{two-level-exact-eigenstates}    
\begin{aligned}
    \dket{0, n} &= \cos\theta_n \ket{0, n} -\sin \theta_n \ket{1, n-1}\,, \\
    \dket{1, n-1} &= \sin \theta_n \ket{0, n} + \cos\theta_n  \ket{1, n-1}
\end{aligned}
\end{equation}
with $\theta_n = \frac{1}{2} \tan^{-1} \left({2\lambda\sqrt{n}}\right)$. In the dispersive limit, we find
\begin{equation}\label{theta_n}
    \theta_n = \lambda\sqrt{n} - \frac{4}{3}\left(\lambda\sqrt{n}\right)^3 + \mathcal{O}\left(\left(\lambda\sqrt{n}\right)^3\right)\,.
\end{equation}
In addition, the dressed state $\dket{0,0}$ is exactly its bare version $\ket{0, 0}$. 

\subsection{Dressed Fock states}

We start by calculating the qubit purity for the exact eigenstate $\dket{0, n}$ given by Eq.~\eqref{two-level-exact-eigenstates}. To this end, we first write the full density matrix:
\begin{multline}\label{rho_0n_full}
\hat{\rho}_{\overline{0, n}} =   \cos^2\theta_n \ketbra{0, n} 
+ \sin^2\theta_n \ketbra{1, n-1} \\- \sin \theta_n \cos\theta_n\left(\ketbra{0, n}{1, n-1} 
+ \ketbra{1, n-1}{0, n}\right)\,.
\end{multline}
The reduced qubit density matrix is then given by
\begin{equation}\label{rho_0n_qubit}
\hat{\rho}^q_{\overline{0, n}} = {\rm Tr}_{\rm res} \hat{\rho}_{\overline{0, n}} = \cos^2\theta_n \ketbra{0} + \sin^2\theta_n \ketbra{1}\,,
\end{equation}
which demonstrates that the qubit is not in a pure state. There are no off-diagonal terms in the reduced density matrix because the terms that are off-diagonal in the full density matrix \eqref{rho_0n_full} are also off-diagonal in the resonator index and do not survive the partial-trace operation. The purity of the reduced density matrix~\eqref{rho_0n_qubit} is given by
\begin{equation}\label{purity_0n}
\mathcal{P}_{{\overline{0, n}}} = {\rm Tr}\left(\hat{\rho}^q_{\overline{0, n}}\right)^2 = \cos^4\theta_n + \sin^4\theta_n\,.
\end{equation}
In the dispersive limit of small $\theta_n$, we observe that $\mathcal{P}_{{\overline{0, n}}} \approx 1 - 2\sin^2\theta_n$. Thus, the qubit purity decreases twice as fast as the probability of finding the qubit in its bare ground state, which is simply $1-\sin^2\theta_n$. 
This cannot explain our numerical observations that purity remains close to $1$ up to relatively large photon numbers during the readout drive.

\subsection{Dressed coherent states}
\label{appendix:dressed_coherent}

We now consider a generic state of the system within the same ground-state eigenladder:
\begin{equation}\label{eq:psi-generic}
\ket{\psi} = \sum_n C_n \dket{0, n}\,,
\end{equation}
where $\sum_n |C_n|^2 = 1$. A dressed coherent state of Eq.~\eqref{eq:dressed} is formed for $C_n = {\alpha^n e^{-|\alpha|^2/2}}/{\sqrt{n!}}$. The full density matrix for the state \eqref{eq:psi-generic} is given by
%\begin{widetext}
\begin{multline}
\hat{\rho}_\psi =  \sum_{n, n'} C_n C^*_{n'} \left(\cos\theta_n \ket{0, n} -\sin \theta_n \ket{1, n-1}\right) 
\\ \times 
\left(\cos\theta_{n'} \bra{0, n'} -\sin \theta_{n'} \bra{1, n'-1}\right)\,.
\end{multline}
%\end{widetext}
To find the reduced qubit density matrix $\hat{\rho}^q_\psi$, we only need to keep track of the bra-ket pairs with the same photon number. Such pairs come from the diagonal terms when $n=n'$ and the off-diagonal terms when $n=n'\pm 1$ (e.g., from the same or nearest-neighbor RWA strips). Thus, we find
\begin{equation}\label{rho_psi_q}
\hat{\rho}^q_\psi
=\rho^q_{00}\ketbra{0} + \rho^q_{11}\ketbra{1} +\rho^q_{01}\ketbra{0}{1} + \left(\rho^{q}_{01}\right)^*\ketbra{1}{0}\,,
\end{equation}
where
\begin{equation}\label{rho_psi_q_00}
\rho^q_{00} = \langle \cos^2\theta_n\rangle = \sum_n |C_n|^2 \cos^2\theta_n\,,
\end{equation} 
\begin{equation}\label{rho_psi_q_11}
\rho^q_{11} = \langle \sin^2\theta_n\rangle=\sum_n |C_n|^2 \sin^2\theta_n\,,
\end{equation} 
and 
\begin{equation}\label{rho_psi_q_01}
\rho^{q}_{01}=-\sum_n C_n C^*_{n+1}\cos\theta_n \sin\theta_{n+1}\,.
\end{equation} 
This more generic expression contrasts with the reduced density matrix given in Eq.~\eqref{rho_0n_qubit} for an eigenstate. While the diagonal matrix elements of $\hat{\rho}^q_\psi$ are simply averages of those in Eq.~\eqref{rho_0n_qubit}, off-diagonal elements appear only in the general expression~\eqref{rho_psi_q}. They arise from the joint contributions of the nearest-neighbor RWA strips and \emph{increase} the qubit purity. For a state $\ket{\psi}$ that is almost evenly ``spread out'' over many RWA strips such as the dressed coherent state $\dket{0, \alpha}$ with $|\alpha|^2\gg 1$, we have $C_n \sim C_{n+1}$ and $\theta_n \sim \theta_{n+1}$ for the dominant terms, so $|\rho^q_{01}| \sim \sqrt{\rho^q_{00}\rho^q_{11}}$. For a perfectly pure state, of course, $|\rho^q_{01}| = \sqrt{\rho^q_{00}\rho^q_{11}}$.

The generalization of Eq.~\eqref{purity_0n} for the qubit purity is given by 
\begin{widetext}
\begin{multline}
    \mathcal{P}_{\psi} = {\rm Tr}\left(\hat{\rho}^q_{\psi}\right)^2 = \left(\sum_n |C_n|^2\cos^2\theta_n\right)^2 + \left(\sum_n |C_n|^2\sin^2\theta_n\right)^2 
    + 2\left|\sum_n C_n C^*_{n+1}\cos\theta_n\sin\theta_{n+1}\right|^2 
    \\
    = 1 -2 \left(\sum_n |C_n|^2\cos^2\theta_n\right)\left(\sum_n |C_n|^2\sin^2\theta_n\right) + 2\left|\sum_n C_n C^*_{n+1}\cos\theta_n\sin\theta_{n+1}\right|^2 \,.
\end{multline}
Since $\theta_n=0$ when $n=0$, we can shift index $n$ by one in the sum containing $\sin^2\theta_n$ to find that
\begin{multline}
    \mathcal{P}_{\psi} = 1 - 2\left[\sum_{n, n'}  \left(|C_n|^2 |C_{n'+1}|^2\cos^2\theta_n \sin^2\theta_{n'+1} - C_n C^*_{n+1} C^*_{n'}C_{n'+1}\cos\theta_n \sin\theta_{n+1} \cos\theta_{n'}\sin\theta_{n'+1}\right)\right] 
    \\
    = 1 - \sum_{n, n'} \left[|C_n|^2 |C_{n'+1}|^2\cos^2\theta_n \sin^2\theta_{n'+1} +|C_{n'}|^2 |C_{n+1}|^2\cos^2\theta_{n'} \sin^2\theta_{n+1} \right. \\
    \left.- (C_n C^*_{n+1} C^*_{n'}C_{n'+1} + C_{n'} C^*_{n'+1} C^*_{n}C_{n+1})\cos\theta_n \sin\theta_{n+1} \cos\theta_{n'}\sin\theta_{n'+1}\right] 
    \\
    = 1-\sum_{n, n'}\left|C_n C_{n'+1}\cos\theta_n \sin\theta_{n'+1} - C_{n'} C_{n+1}\cos\theta_{n'} \sin\theta_{n+1}\right|^2 
\end{multline}
For a dressed coherent state $\dket{0, \alpha}$, we thus have
\begin{equation}\label{purity_0alpha_start}
    \mathcal{P}_{\overline{0, \alpha}}= 1 - e^{-2|\alpha|^2} \sum_{n, n'}\frac{|\alpha|^{2(n+ n'+1)}}{n! n'!} \left(\frac{\cos\theta_n \sin\theta_{n'+1}}{\sqrt{n'+1}} - \frac{\cos\theta_{n'}\sin\theta_{n+1}}{\sqrt{n+1}}\right)^2\,.
\end{equation}

\subsection{Dispersive limit}
\label{appendix:purity_dispersive_limit}

Let us find the leading contribution in the dispersive limit when $|\alpha|^2 \ll \Delta^2/(4g^2)$. We first notice that if we simply use $\theta_n \approx \lambda \sqrt{n}$ and $\cos\theta_n \approx 1$, the result would be $\mathcal{P}_{\overline{0, \alpha}}=1$, so we should consider higher-order corrections. Using Eq.~\eqref{theta_n} and keeping the terms up to the cubic power in $\lambda\sqrt{n}$, we find
\begin{equation}
\begin{aligned}
\sin\theta_n &=  \lambda\sqrt{n} - \frac{3}{2} (\lambda\sqrt{n})^3 + \mathcal{O}\left((\lambda\sqrt{n})^3\right)\,, \\
\cos\theta_n &=  1 - \frac 12 (\lambda\sqrt{n})^2 + \mathcal{O}\left((\lambda\sqrt{n})^3\right)\,,
\end{aligned}
\end{equation}
and
\begin{equation}
\frac{\cos\theta_n \sin\theta_{n'+1}}{\sqrt{n'+1}} =  \lambda - \left[\frac{n}{2} + \frac{3(n'+1)}{2}\right]\lambda^3 + \mathcal{O}(\lambda^3)\,,
\end{equation}
so
\begin{equation}
\left(\frac{\cos\theta_n \sin\theta_{n'+1}}{\sqrt{n'+1}} - \frac{\cos\theta_{n'}\sin\theta_{n+1}}{\sqrt{n+1}}\right)^2 
= (n-n')^2 \lambda^6 + \mathcal{O}(\lambda^6)\,.
\end{equation}
Therefore, the effect of the qubit purity being different from one is in the sixth order in $g/\Delta$! In reality, deviations of the real state from the dressed coherent state and deviations from the dispersive limit can give lower-order terms in~$\lambda$. 

To evaluate Eq.~\eqref{purity_0alpha_start}, we calculate
\begin{multline}
f(\alpha) = e^{-2|\alpha|^2} \sum_{n, n'}\frac{|\alpha|^{2(n+ n'+1)}}{n! n'!} (n-n')^2 
= 2e^{-|\alpha|^2}\sum_n \frac{|\alpha|^{2(n+1)}}{n!}n^2 - 2e^{-2|\alpha|^2} \left(\sum_n \frac{|\alpha|^{2n + 1}}{n!}n\right)^2 
\\
= 2e^{-|\alpha|^2}\sum_{n\ge 1} \frac{|\alpha|^{2(n+1)}}{(n-1)!}[(n-1) + 1] - 2e^{-2|\alpha|^2} \left(\sum_{n\ge 1} \frac{|\alpha|^{2n + 1}}{(n-1)!}\right)^2 
 = 2|\alpha|^4\,.
\end{multline}
Therefore, we find in the dispersive approximation
\begin{equation}\label{purity_0alpha_dispersive}
    \mathcal{P}_{\overline{0, \alpha}} =1 - 2|\alpha|^4 \left(\frac{g}{\Delta}\right)^6\,.
\end{equation}
In Fig.~\ref{fig:purity}, we show this approximate result by dashed lines.

\subsection{Strong hybridization}
\label{appendix:strong_hyridization}

Let us now consider the regime opposite to the dispersive approximation. Namely, we assume that $|\alpha|^2 \gg \Delta^2/g^2$, so the Fock states hybridize very strongly and $\theta_n \sim \pi/4$ for relevant $n$. Using $\cos\theta_n = \sin\theta_n = 1/2$ in Eq.~\eqref{purity_0alpha_start}, we find
\begin{multline}
      \mathcal{P}_{\overline{0, \alpha}} =  1 - \frac{e^{-2|\alpha|^2}}{16} \sum_{n, n'}\frac{|\alpha|^{2(n+ n'+1)}}{n! n'!} \left(\frac{1}{\sqrt{n'+1}} - \frac{1}{\sqrt{n+1}}\right)^2 
      = 1 
      - \frac{e^{-2|\alpha|^2}}{8} \sum_{n,n'} \frac{|\alpha|^{2n} |\alpha|^{2(n'+1)}}{n! (n'+1)!} 
      \\+ \frac{e^{-2|\alpha|^2}}{8} \left(\sum_{n}\frac{|\alpha|^{2n+1}}{n!\sqrt{n+1}} \right)^2 
      = 1 - \frac{1-e^{-|\alpha|^2}}{8} + \frac{1}{8} \left[1 + e^{-|\alpha|^2}\sum_n\frac{|\alpha|^{2n}(|\alpha|-\sqrt{n+1})}{n!\sqrt{n+1}}\right]^2\,.
\end{multline}
This expression suggests that even in this strong-coupling regime, the dressed coherent state is very close to being pure.

\end{widetext}

\section{Dissipation effects}
\label{appendix:dissipation}

\begin{figure}
    \centering
    \includegraphics[width=\columnwidth]{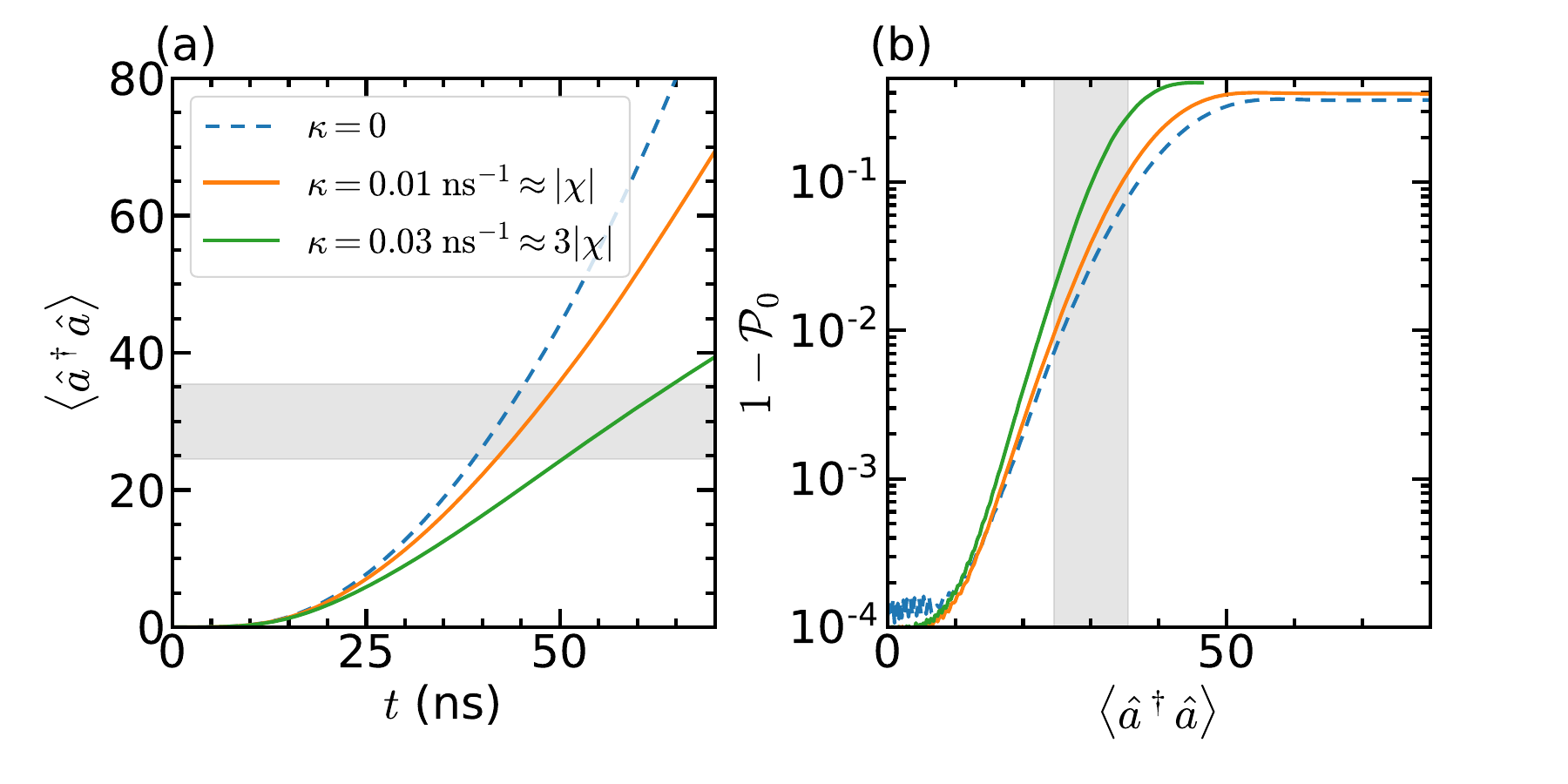}
    \vspace{-20pt}
    \caption{Time evolution of the initial state $\dket{0, 0}$ in the presence of the resonant readout drive with $\varepsilon/2\pi = 25\,\mathrm{MHz}$ for a~coherent dynamics (dashed lines) and in the presence of dissipation (solid lines) for parameters of Fig.~\ref{fig:metrics}. (a) Resonator photon occupation $\expval{\hat{a}^\dagger \hat{a}}$ versus time~$t$. (b) Purity error $1-\mathcal{P}_0$ versus $\expval{\hat{a}^\dagger \hat{a}}$ with the dashed line matching the dashed line of Fig.~\ref{fig:metrics}(c). Gray shading highlights the same regions as in Fig.~\ref{fig:metrics}.}
    \label{fig:dissipation}
\end{figure}

In this appendix, we verify the robustness of qubit purity to the dissipation effects. To this end, we simulate the dissipative dynamics for parameters of Fig.~\ref{fig:metrics} by integrating the Lindblad master equation with the collapse operator $\sqrt{\kappa}\hat{a}$ using the Monte-Carlo-quantum-trajectories solver of QuTiP~\cite{Johansson2012, Johansson2013}. In Fig.~\ref{fig:dissipation}, we show the results of these simulations for $\varepsilon/2\pi = 25\,\mathrm{MHz}$ and two different values of the resonator decay rate $\kappa$, which are close to $|\chi|$ and $3|\chi|$ with $\chi/2\pi \approx -1.55\,\mathrm{MHz}$ being the resonator dispersive shift for given parameters. First, we find that the purity error $1-\mathcal{P}_0$ remains very close to its value for a coherent system (dashed line). Second, comparing Figs.~\ref{fig:metrics}(c) and \ref{fig:dissipation}(b), we observe that the effect of $\kappa$ is opposite to that of $\varepsilon$ since both increasing $\kappa$ and reducing $\varepsilon$ result in a higher probability of an adiabatic transition at an avoided-level-like crossing.

\section{Perturbation theory for matrix elements}
\label{appendix:perturbation_mat_el}

Here, we calculate the matrix elements of $\hat{a}^\dagger$ for the states within the same eigenladder perturbatively, assuming charge coupling. For the same eigenladder, the first-order perturbation theory gives zero correction, and we need to perform the calculations in the second order.
For fluxonium, the perturbation (qubit-resonator interaction) has the form
\begin{equation}\label{qubres_coupling_Hamiltonian}
\hat{V} \equiv \hat{V}_{qr}(g) = -i\hbar g\hat{n}\left(\hat{a} - \hat{a}^\dagger\right)\,.
\end{equation}
\begin{widetext}
\noindent For the bare states, we have 
\begin{equation}\label{V_k'n'_kn}
\bra{k', n'} \hat{V} \ket{k, n} = ig O_{k', k} \left( \sqrt{n+1}\delta_{n', n+1} - \sqrt{n}\delta_{n', n-1}\right)\,,
\end{equation}
where
$O_{k', k} = \hbar \bra{k'} \hat{n}\ket{k}$\,.
Then, the dressed state up to the second order in $g$, including the normalization correction, has the form

\begin{multline}
\dket{k, n} = \left[1-\frac 12 \sum_{\{k'n'\}\ne \{kn\}} \frac{|\bra{k', n'} \hat{V} \ket{k, n}|^2}{\left(E_{k,n} - E_{k',n'}\right)^2}\right]\ket{k, n} + \sum_{\{k',n'\}\ne \{k,n\}} \frac{\bra{k', n'} \hat{V} \ket{k, n} }{E_{k,n} - E_{k',n'}} \ket{k',n'} 
\\
+ \sum_{\{k',n'\}, \{k'',n''\}\ne \{k,n\}}  \frac{\bra{k', n'} \hat{V} \ket{k'', n''} \bra{k'', n''} \hat{V} \ket{k, n} }{\left(E_{k,n} - E_{k',n'}\right) \left(E_{k,n} - E_{k'',n''}\right)} \ket{k',n'} 
- \sum_{\{k',n'\}\ne \{k,n\}} \frac{\bra{k, n} \hat{V} \ket{k, n} \bra{k', n'} \hat{V} \ket{k, n}}{\left(E_{k,n} - E_{k',n'}\right)^2} \ket{k',n'} \,.
\end{multline}
The last term in this expression is exactly zero because it contains a diagonal matrix element of $\hat{V}$. The other terms yield
\begin{multline}
\dket{k, n} = \left\{ 1 - \frac{g^2}{2}\sum_{k'}|O_{k', k}|^2\left[\frac{n+1}{(E_k - E_{k'}-\omega_r)^2} + \frac{n}{(E_k - E_{k'}+\omega_r)^2} \right]\right\}\ket{k, n} 
\\
+  i g\sum_{k'} O_{k', k}\left[\frac{\sqrt{n+1}}{E_k - E_{k'} - \omega_r} \ket{k', n+1} - \frac{\sqrt{n}}{E_k - E_{k'} + \omega_r} \ket{k', n-1} \right]
\\
- g^2 \sum_{k', k''} O_{k', k''} O_{k'', k}\left[\frac{\sqrt{(n+2)(n+1)}}{(E_k - E_{k'} - 2\omega_r)(E_k - E_{k''} - \omega_r)} \ket{k', n+2} + \frac{\sqrt{n(n-1)}}{(E_k - E_{k'} + 2\omega_r)(E_k - E_{k''} + \omega_r)} \ket{k', n-2}\right]
\\
+ g^2 \sum_{k'\neq k, k''} O_{k', k''} O_{k'', k}\left[\frac{n}{(E_k - E_{k'})(E_k - E_{k''} + \omega_r)} + \frac{n+1}{(E_k - E_{k'})(E_k - E_{k''} - \omega_r)}\right]\ket{k', n}\,.
\end{multline}
Thus, we find
\begin{multline}
\mel{\overline{k, n'}}{\hat{a}^\dagger}{\overline{k, n}} = \sqrt{n+1} \left\{ 1 - \frac{g^2}{2}\sum_{k'}|O_{k', k}|^2\left[\frac{2n+3}{(E_k - E_{k'}-\omega_r)^2} + \frac{2n+1}{(E_k - E_{k'}+\omega_r)^2} \right]\right\}\delta_{n', n+1} 
\\
+ig O_{k, k} \left[\frac{\sqrt{(n+2)(n+1)}}{-\omega_r}\delta_{n', n+2} -\frac{n}{\omega_r}\delta_{n', n} - \frac{n+1}{-\omega_r} \delta_{n', n} + \frac{\sqrt{(n+2)(n+1)}}{\omega_r}\delta_{n', n+2}\right]
\\
+ g^2 \sum_{k'} |O_{k, k'}|^2 \left\{\left[\frac{(n+2)\sqrt{n+1}}{(E_k - E_{k'}-\omega_r)^2} + \frac{n\sqrt{n+1}}{(E_k - E_{k'}+\omega_r)^2}\right]\delta_{n', n+1} - \frac{n\sqrt{n}\delta_{n', n-1} + \sqrt{(n+3)(n+2)(n+1)}\delta_{n', n+3}}{ (E_k - E_{k'}-\omega_r)(E_k - E_{k'}+\omega_r)}\right\}
\\
- g^2 \sum_{k'} |O_{k, k'}|^2\left[\frac{\sqrt{(n+3)(n+2)(n+1)}}{( - 2\omega_r)(E_k - E_{k'} - \omega_r)} \delta_{n', n+3} + \frac{\sqrt{n}(n-1)}{2\omega_r(E_k - E_{k'} + \omega_r)} \delta_{n', n-1}\right]
\\
- g^2 \sum_{k'} |O_{k, k'}|^2\left[\frac{(n+1)\sqrt{n}}{( - 2\omega_r)(E_k - E_{k'} - \omega_r)} \delta_{n', n-1} + \frac{\sqrt{(n+3)(n+2)(n+1)}}{2\omega_r(E_k - E_{k'} + \omega_r)} \delta_{n', n+3}\right]\,.
\end{multline}
We note that $O_{k, k} = 0$ is the expectation value of the momentum operator in a one-dimensional bound state and that the terms with $\delta_{n', n+3}$ cancel each other out.
Therefore, only the following matrix elements within the same eigenladder are nonzero in second order:
\begin{equation}\label{a_n+1_n_pert}
    \mel{\overline{k, n+1}}{\hat{a}^\dagger}{\overline{k, n}} = \sqrt{n+1}\left\{1 + \frac{g^2}{2}\sum_{k'} |O_{k, k'}|^2 \left[\frac{1}{(E_k - E_{k'}-\omega_r)^2}  - \frac{1}{(E_k - E_{k'}+\omega_r)^2}\right]  \right\}\,
\end{equation}
and
\begin{equation}\label{a_n-1_n_pert}
    \mel{\overline{k, n-1}}{\hat{a}^\dagger}{\overline{k, n}} = 
    g^2\sqrt{n} \sum_{k'} |O_{kk'}|^2 \frac{E_k - E_{k'}}{\omega_r(E_k - E_{k'} - \omega_r) (E_k - E_{k'} + \omega_r)}\,.
\end{equation}

We next find the expectation value for the dressed coherent state:
\begin{multline}\label{a_alpha_alpha_pert}
\mel{\overline{k, \alpha}}{\hat{a}^\dagger}{\overline{k, \alpha}}= 
e^{-|\alpha|^2} \sum_n \frac{|\alpha|^{2n}}{n!\sqrt{n+1}} \left[ \alpha^*\mel{\overline{k, n+1}}{\hat{a}^\dagger}{\overline{k, n}} + \alpha\mel{\overline{k, n}}{\hat{a}^\dagger}{\overline{k, n+1}}\right]
\\
= \alpha^* + \frac{g^2 e^{-|\alpha|^2}}{2} \sum_{k', n} |O_{k, k'}|^2 \frac{|\alpha|^{2n}}{n!} \left[\frac{\alpha^*}{(E_k - E_{k'}-\omega_r)^2}  - \frac{\alpha^*}{(E_k - E_{k'}+\omega_r)^2} + \frac{2\alpha (E_k - E_{k'})}{\omega_r(E_k - E_{k'} - \omega_r) (E_k - E_{k'} + \omega_r)}\right]\,.
\end{multline}
At $\alpha\rightarrow 0$, this expectation value does not simply reduce to Eq.~\eqref{a_n+1_n_pert} but has a correction coming from Eq.~\eqref{a_n-1_n_pert}.
Finally, the matrix-elements error~\eqref{eq:error_mat_el} is given by
\begin{equation}\label{eq:error_mel_pert}
    \mathcal{E}_{k, \alpha} = \frac{g^2}{2} \left|\sum_{k'} |O_{k, k'}|^2  \left[\frac{1}{(E_k - E_{k'}-\omega_r)^2}  - \frac{1}{(E_k - E_{k'}+\omega_r)^2} + \frac{2 (E_k - E_{k'}) (\alpha/\alpha^*)}{\omega_r(E_k - E_{k'} - \omega_r) (E_k - E_{k'} + \omega_r)}\right]\right|,
\end{equation}
which is shown by the dashed line in Fig.~\ref{fig:metrics}(d).
We thus find that $\mathcal{E}_{k, \alpha}$ calculated perturbatively depends on the phase of $\alpha$ but is independent of its magnitude. In simulations in this paper, we used $\alpha = |\alpha|e^{i\pi/2}$.

\end{widetext}

\section{Identifying measurement-induced state transitions with different metrics}
\label{appendix:comparison}

\begin{figure}
    \centering
    \includegraphics[width=\columnwidth]{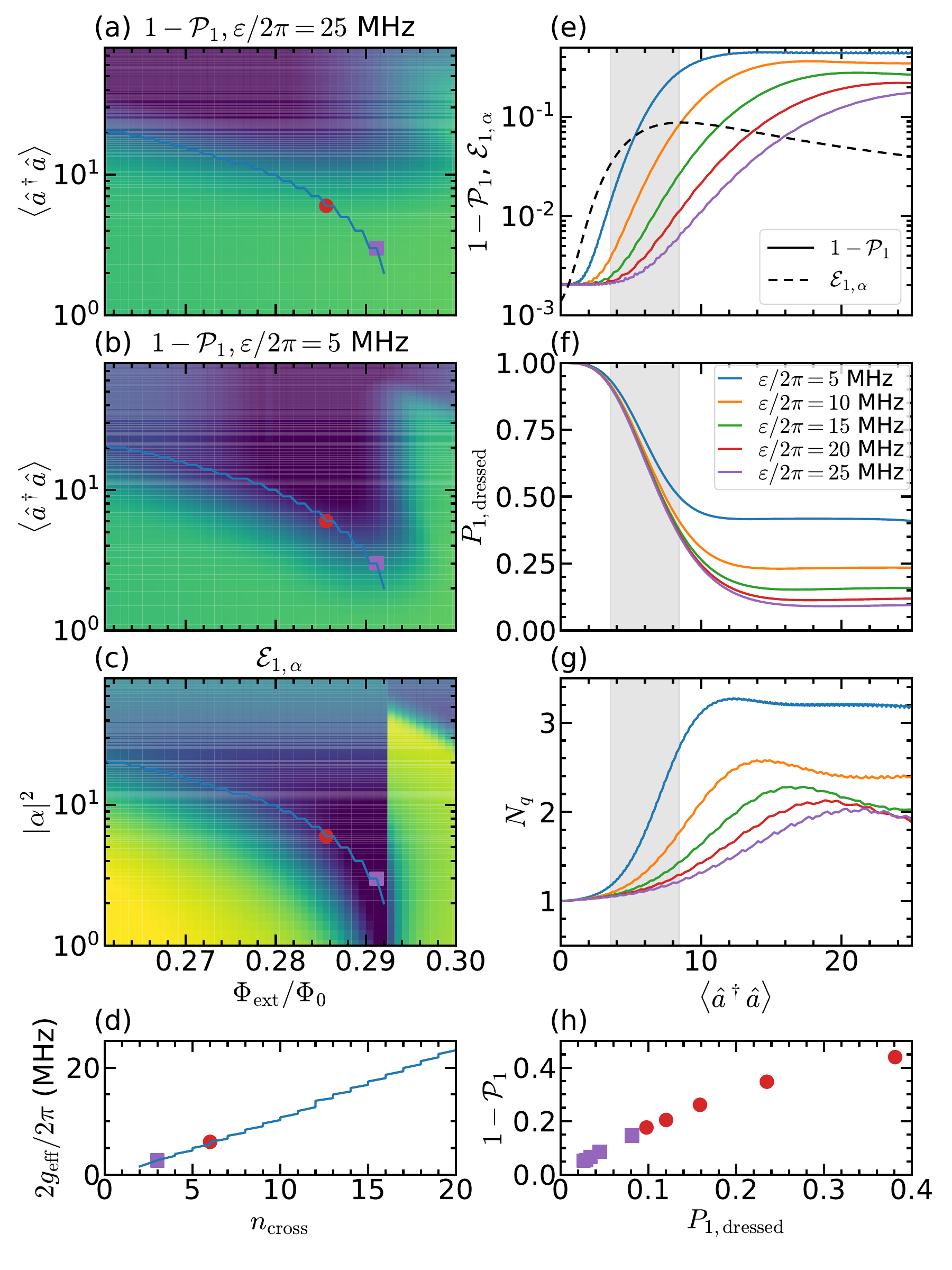}
    \caption{(a) Zoom-in into Fig.~\ref{fig:phi_ext}(f) near $\Phi_{\rm ext}/\Phi_0 = 0.3$. (b) Same but for the purity error $1-\mathcal{P}_1$ calculated for $\varepsilon/2\pi = 5\,\mathrm{MHz}$ instead of $25\,\mathrm{MHz}$. (c) Zoom-in into Fig.~\ref{fig:phi_ext}(h) over the same external flux range. The color scales in panels (a--c) match those in Figs.~\ref{fig:phi_ext}(f) and~\ref{fig:phi_ext}(h). The solid lines in panels (a--c) show $n_{\rm cross}$ defined as the location of the avoided-level-like crossings between eigenenergies $E_{\overline{1, n}}$ and $E_{\overline{6, n-2}}$ with an energy gap $2g_{\rm eff}$ in an analogy to Fig.~\ref{fig:levels}(d). (d)~The parametric plot $2g_{\rm eff}$ versus $n_{\rm cross}$. Circles and squares highlight identical flux values in the left-column panels. In panels (e--g), solid lines show $1-\mathcal{P}_1$~(e), $P_{\rm 1, dressed}$~(f), and qubit occupation $N_q\equiv\sum_{k}k\expval{\hat{\rho}_q}{k}$~(g) versus $\expval{\hat{a}^\dagger \hat{a}}$ for 5 values of the drive amplitude~$\varepsilon$ and the flux corresponding to the red circles in the left column. The area around $n_{\rm cross}=6$ is shaded; and the additional dashed line in panel (e) shows $\mathcal{E}_{1, \alpha}$ versus $|\alpha|^2$. (h) Squares (circles): purity error $1-\mathcal{P}_1$ versus $P_{1, \rm dressed}$ calculated at $\expval{\hat{a}^\dagger \hat{a}} = 25\,(30)$ for the left-column purple squares (red circles) and for the same set of $\varepsilon$ values as in panels (e--g). 
    }
    \label{fig:zoomin}
\end{figure}

In this appendix, we investigate the dependence of qubit purity on the drive amplitude and compare different metrics in more detail. To this end, we focus on the features near $\Phi_{\rm ext}/\Phi_0=0.3$ in the right column of Fig.~\ref{fig:phi_ext}, which correspond to the two-photon resonance condition $\omega_{16} \approx 2\omega_r$, see Fig.~\ref{fig:phi_ext}(b). In Figs.~\ref{fig:zoomin}(a) and \ref{fig:zoomin}(c), we show zoom-ins into Figs.~\ref{fig:phi_ext}(f) and \ref{fig:phi_ext}(h), while Figure~\ref{fig:zoomin}(b) shows the purity error recalculated for a much smaller drive amplitude $\varepsilon/2\pi = 5\,\mathrm{MHz}$ but for a longer pulse time of 400\,ns. [For visual clarity, the data in Fig.~\ref{fig:zoomin}(b) are extrapolated whenever the maximum resonator occupation reached is below the vertical axis limit of 80 photons.] We immediately observe that when compared to Fig.~\ref{fig:zoomin}(a), the onset of features in $1-\mathcal{P}_1$ in Fig.~\ref{fig:zoomin}(b) generally occurs at a~lower resonator occupation $\bar{n}$, with the value closer to the onset of the feature in $\mathcal{E}_{1, \alpha}$ seen in Fig.~\ref{fig:zoomin}(c). To understand this effect, we examined the avoided-level-like crossings between dressed energies of states $\dket{1, n}$ and $\dket{6, n-2}$ at various $\Phi_{\rm ext}$ by calculating their locations $n_{\rm cross}$ and the corresponding energy gaps $2g_{\rm eff}$, following the methodology of Fig.~\ref{fig:levels}(d). We superimpose $n_{\rm cross}$ versus $\Phi_{\rm ext}$ in all the heat maps of Fig.~\ref{fig:zoomin}, and we plot $2g_{\rm eff}$ versus $n_{\rm cross}$ in Fig.~\ref{fig:zoomin}(d).

Figure~\ref{fig:zoomin}(d) reveals that the gap $2g_{\rm eff}$ grows with $n_{\rm cross}$ monotonically. Moreover, because of the two-photon-resonance condition, this dependence is almost linear as expected for a second-order effect with the matrix elements of the qubit-resonator interaction $\hat{V}_{qr}(g)$ scaling as $g\sqrt{n}$. We next observe that the onset of the features in $1-\mathcal{P}_1$ is less sensitive to the drive amplitude $\varepsilon$ when $n_{\rm cross}$ and therefore the gap $2g_{\rm eff}$ are both larger. That is, this onset occurs at almost the same $\bar{n}\sim n_{\rm cross}$ on all three heat maps at $\Phi_{\rm ext}/\Phi_0 \lesssim 0.27$, where $2g_{\rm eff}/2\pi$ approaches 20\,MHz, causing increased adiabaticity of the transition through the avoided-level-like crossing during the resonator ring-up. In contrast, for the flux corresponding to the purple square, where $2g_{\rm eff}/2\pi$ is only 3\,MHz and the adiabaticity is noticeably reduced for large~$\varepsilon$, the onset for $\varepsilon/2\pi=25\,\mathrm{MHz}$ occurs at $\bar{n}$ several times larger than $n_{\rm cross}=3$. As a result, the value of $\bar{n}$ for the onset of features is higher for $1-\mathcal{P}_1$ calculated at $\varepsilon/2\pi=25\,\mathrm{MHz}$ than for $\mathcal{E}_{1, \alpha}$. A similar observation holds for other multiphoton resonances in Figs.~\ref{fig:phi_ext} and \ref{fig:omega_r}.

In Fig.~\ref{fig:zoomin}(e), we further explore the sensitivity of the purity error to the degree of adiabaticity of traversal through the avoided-level-crossing-like region. We show $1-\mathcal{P}_1$ versus $\expval{\hat{a}^\dagger \hat{a}}$ calculated for several values of $\varepsilon$ (solid lines) at the flux corresponding to the red circles in the left column, when $n_{\rm cross}=6$ and $2g_{\rm eff}/2\pi \approx 6\,\mathrm{MHz}$. We supplement these curves by showing $\mathcal{E}_{1, \alpha}$ versus $|\alpha|^2$ on the same plot (dashed line) and $P_{1, \rm dressed}$ and qubit occupation $N_q = \sum_k k \langle k|\hat{\rho}_q|k\rangle$ in Figs.~\ref{fig:zoomin}(f) and \ref{fig:zoomin}(g) for the same set of $\varepsilon$. For the smallest $\varepsilon/2\pi=5\,\mathrm{MHz}$ (blue lines), the error $1-\mathcal{P}_1$ is accumulated mainly within the avoided-level-crossing-like region (shaded area), with a~rapid increase appearing closer to the one in the drive-independent metric $\mathcal{E}_{1, \alpha}$. In this case, the probability of the adiabatic transition and therefore of staying within the $\dket{1, n}$ eigenladder is near 40\%, resulting in an approximately 60\%-40\% population split between the bare qubit levels $\ket{1}$ and $\ket{6}$ and $N_q\sim 3$. We explain the earlier onset of changes in both $1-\mathcal{P}_1$ at smaller $\varepsilon$ and $\mathcal{E}_{1, \alpha}$ by the metrics sensitivities to the right tail of the photon-number distribution in $\dket{1, \alpha}$ entering the avoided-level-crossing-like region ahead of the average reaching $n_{\rm cross} - \sqrt{n_{\rm cross}}$. In comparison, for the largest $\varepsilon/2\pi=25\,\mathrm{MHz}$ (purple lines), when the probability of the adiabatic transition is below 10\% and about half of the change in $N_q$ comes from the increased hybridization of the bare states in $\dket{1, n}$ rather than the MIST, $1-\mathcal{P}_1$ is accumulated mainly after the avoided-level-crossing-like region. In this case, the accumulation of the purity error is due to the left tail of the photon-number distribution in the state $\dket{1, \alpha}$. During the resonator ring-up, the width of the distribution increases with the points in its left tail moving more slowly than those in its center and right tail, increasing the relative contribution of the left tail to the probability of the adiabatic transition.

Finally, we briefly discuss the relation between $1-\mathcal{P}_1$ and the MIST probability, which is given by the probability of the adiabatic transition, i.e., $P_{1, \rm {dressed}}$ after the avoided-level-crossing-like region; see the end of Sec.~\ref{Sec-metrics-probabilities}. In Fig.~\ref{fig:zoomin}(h), we show $1-\mathcal{P}_1$ versus $P_{1, \rm {dressed}}$ with the values taken after the avoided-level-crossing-like region, when both metrics saturate. Here, the points are sampled from the simulations for~$\Phi_{\rm ext}$ marked by circles and squares in the left column of Fig.~\ref{fig:zoomin}. We find that the purity error increases monotonically with the MIST probability. We note that this observation is not general and may be violated when $2g_{\rm eff}$ is larger and the adiabaticity is higher. Nevertheless, Fig.~\ref{fig:zoomin}(h) suggests that the purity error $1-\mathcal{P}_k$ is a good proxy metric when the probability of MISTs is small, an experimentally relevant high-fidelity readout regime.

\section{Measurement-induced state transitions in a transmon qubit}
\label{appendix:transmon}

\begin{figure}
    \centering
    \includegraphics[width=\columnwidth]{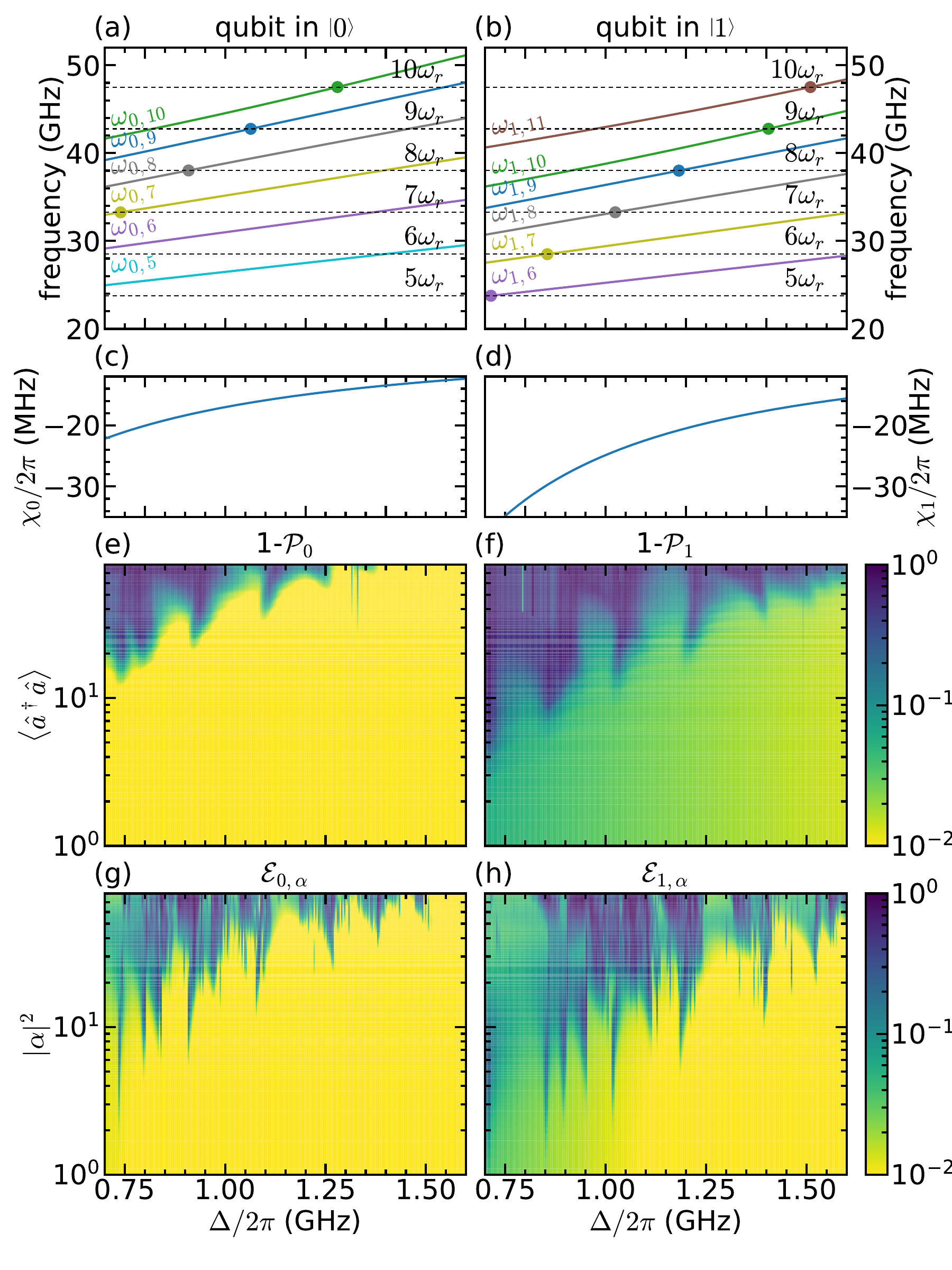}
    \caption{MISTs in a transmon as a function of the qubit-resonator detuning $\Delta=\omega_{0,1}-\omega_r$ (horizontal axes). The notation and the panel arrangement are the same as in Figs.~\ref{fig:phi_ext} and~\ref{fig:omega_r} of the main text. The simulations are done for parameters of Ref.~\cite{Khezri23} but for a fixed $n_g=0.2$ instead of averaging over its full range so that the transition-frequency diagrams in panels (a) and (b) are not smeared out. Round markers in panels~(a) and (b) label only those resonances for which state transitions are compatible with the selection rules of Eq.~\eqref{eq:Vqr-transmon}.}
    \label{fig:transmon}
\end{figure}

To facilitate comparison with a recent transmon study~\cite{Khezri23}, we include the simulations for a transmon qubit defined by the Hamiltonian~\cite{Koch07}
\begin{equation}\label{eq:transmon}
\hat{H}_q = 4E_C (\hat{n}-n_g)^2 - E_J \cos\hat{\varphi}.
\end{equation}
Here, $n_g$ is the offset charge, which can often be ignored in the transmon regime $E_J/E_C \gg 1$ but must be taken into account in a MIST study due to the sensitivity of higher-lying transmon levels to $n_g$~\cite{Khezri23}.
We intentionally do not assume any approximation and use the drive term of the form~\eqref{eq:drive} together with the qubit-resonator interaction term
\begin{equation}\label{eq:Vqr-transmon}\nonumber
        \hat{V}_{qr}(g) = i\hbar g (\hat{n}-n_g)(\hat{a}^\dagger - \hat{a})\,.
        \end{equation}
We present our simulations in Fig.~\ref{fig:transmon} for the parameters of Ref.~\cite{Khezri23}: $\omega_r/2\pi = 4.751\,\textrm{GHz}$, $E_C/h=0.194\,\textrm{GHz}$, $\varepsilon/2\pi=45\,\textrm{MHz}$, and $k_{\rm eff} = 0.048$ for the qubit-resonator coupling efficiency. For each value of the qubit-resonator detuning $\Delta = \omega_{01} - \omega_r$, varied in the range between 0.7 and 1.6\,GHz, we calculate $E_J$ to match the qubit frequency~$\omega_{01}$ and find the coupling strength $g = k_{\rm eff}  \sqrt{\omega_{01}\omega_r}/(2{\mel{0}{\hat{n}}{1}})\approx 2\pi \times 92.2\,\textrm{MHz}$, which is independent of $E_J$. Although the averaging over $n_g$ is straightforward and is necessary to explain the experimental observations in Ref.~\cite{Khezri23}, the simulations in Fig.~\ref{fig:transmon} are done for a fixed $n_g=0.2$ to preserve the clarity of the transition-frequency diagrams in Figs.~\ref{fig:transmon}(a) and~\ref{fig:transmon}(b). Even with a~fixed $n_g$, the simulations agree very well with the experimental data of Ref.~\cite{Khezri23}. These simulations do not rely on the RWA approximation used in Ref.~\cite{Khezri23}, although the RWA approximation can provide similar results in a~faster computational runtime.

\bibliography{literature}

\end{document}